\documentclass[preprint,12pt]{elsarticle}
\usepackage{graphicx}
\usepackage{dcolumn}
\usepackage{bm}

\usepackage[utf8]{inputenc}
\usepackage[T1]{fontenc}
\usepackage{amsmath}
\usepackage{booktabs}
\usepackage{multirow}
\usepackage{mathptmx}
\usepackage{etoolbox}
\usepackage{subcaption}
\usepackage{xcolor}
\usepackage{adjustbox} 
\usepackage{rotating}  
\graphicspath{{./figuresEPS/}}
\usepackage{a4wide}
\usepackage{hyperref}
\usepackage{cleveref}

\usepackage{booktabs} 

\journal{IJHFF}
\begin{document}

\newcommand{\Revone}[1]{\textcolor{black}{#1}}

\begin{frontmatter}

\title{\centering Effect of Chordwise Flexibility Distribution on Wave-Assisted Flapping Foil Performance}

\author[a]{Lokesh Silwal}
\author[b]{Neel Karani}
\author[a,b]{Anchal Sareen\textsuperscript{\textdagger}}

\affiliation[a]{Department of Naval Architecture and Marine Engineering, University of Michigan, Ann Arbor, MI 48109, USA}
\affiliation[b]{Department of Mechanical Engineering, University of Michigan, Ann Arbor, MI 48109, USA}
\cortext[cor1]{Corresponding author: Anchal Sareen, Email: asareen@umich.edu}


\begin{abstract}

This study investigates the influence of the spatial distribution of flexibility along a propulsor on thrust generation and propulsive efficiency in wave-assisted flapping foils. While flexibility is known to enhance propulsive performance, the role of its chordwise placement remains poorly understood. Here, the effective flexible length is systematically varied by shifting the flexure location along the tail while maintaining constant flexural rigidity and total chord length. Experiments are conducted in quiescent flow at heave frequencies of $0.8~\mathrm{Hz}$ and $1.25~\mathrm{Hz}$, and non-dimensional heave amplitudes of $h^* = 0.13$ and $0.22$. Simultaneous measurements of hydrodynamic forces, flow fields, and tail kinematics are used to quantify performance and elucidate the underlying fluid–structure interactions. The fully flexible configuration consistently achieves higher propulsive efficiency (up to $\sim 164\%$) across all conditions, which is attributed to enhanced jet persistence and increased streamwise vortex spacing, indicative of a more coherent and sustained momentum jet. In contrast, the mid-flexible configuration yields substantially higher thrust (up to $\sim 66\%$) at the largest heave frequency and amplitude, driven by a pronounced increase in near-wake jet velocity and momentum flux. These results demonstrate that the chordwise distribution of flexibility governs the trade-off between thrust and propulsive efficiency by modulating wake coherence and momentum transfer. The findings establish flexibility placement as a key design parameter in flapping propulsion and provide physics-based guidelines for enhancing the performance and endurance of wave-driven unmanned surface vehicles.

\end{abstract}






\end{frontmatter}


\section{\label{sec:intro}Introduction}

The increasing urgency of climate change has motivated the development of carbon-neutral marine propulsion technologies. \Revone{Among emerging approaches, wave-assisted propulsion (WAP) has attracted significant attention due to its potential to harvest ambient wave energy for thrust generation, offering a compelling engineering pathway toward reduced-emission unmanned surface vehicles (USVs). From a design standpoint, WAP systems present a tractable fluid-structure interaction problem with direct implications for propulsor selection, structural compliance, and endurance optimization in engineered marine systems.} WAP systems are inspired by bio hydrodynamic mechanisms observed in aquatic locomotion, where many species generate thrust through combined heaving and pitching motions of caudal fins \citep{paraz_thrust_2016}. These principles can be adapted for surface vessels by elastically mounting a hydrofoil to the hull of the vehicle enabling flapping motion as the vehicle heaves up and down with the surface waves. Such systems have been shown to exhibit functional similarities to conventional rotary propulsors \citep{thaweewat2018semi}.
Most existing WAP configurations employ passively pitching foils driven by wave-induced heave, typically using torsional springs to enable pitch motion. This design choice reduces actuation power requirements and has been shown to yield higher propulsive efficiencies compared to actively pitch-controlled foils \citep{bockmann2014experiments}. The influence of key parameters, including torsional stiffness, reduced frequency, foil mass, and pivot location, has been systematically investigated through numerical and experimental studies \citep{qi2019numerical,qi2020effect,yang2018numerical,yang2019systematic}. More recently, Raut et al. \citep{raut2024hydrodynamic} examined the effects of spring stiffness and pivot location using a simplified model of a single, sinusoidally heaving foil with passive pitch. This study identified the leading-edge vortex as the primary mechanism governing thrust production. Building on these findings, subsequent investigations extended the analysis to tandem hydrofoil configurations \citep{raut2025dynamics} and examined the role of pitch limiters in shaping the propulsive performance \citep{Raut2025}.\\

While most existing WAP studies have focused on passive pitching of rigid foils, chordwise flexibility offers an alternative pathway for enhancing propulsive performance. The geometric and material parameters that define flexible propulsors, including foil shape, compliance distribution, and trailing-edge configuration, are known to govern wake dynamics and force generation in ways that are directly relevant to the design of engineered propulsion systems.Surface topology provides an additional means of modifying hydrodynamic forces and wake dynamics without changing the underlying body geometry. Recent studies have shown that surface dimples can reorganize vortex shedding and wake coherence on both fixed and pitching foils, with effects that depend strongly on the Reynolds number and operating condition \cite{Sudarsana2025,Silwal2026}. More broadly, dynamically morphable roughness has been used to achieve adaptive drag reduction and generate substantial lateral forces on bluff bodies \cite{Vilumbrales-Garcia2025,Sudarsana2024}. These findings establish surface morphology as a potentially programmable hydrodynamic design variable complementary to structural flexibility. For instance, Kelly \citep{kelly2023geometric} demonstrated through numerical simulation that the chordwise thickness distribution of a flapping foil governs vortex formation and the interference between leading- and trailing-edge vortices, with performance showing strong sensitivity to the foil geometry in the 18 to 50\% chord region. Flexibility may be introduced either through distributed chordwise compliance of the foil or by appending a flexible tail to an otherwise rigid foil. Both approaches have been shown to improve thrust generation in flapping-foil propulsion. For example, Marais et al.\citep{marais2012stabilizing} investigated a teardrop-shaped foil with chordwise flexibility and reported up to a threefold increase in thrust relative to a rigid foil. In addition to enhancing thrust, flexibility was shown to suppress symmetry breaking of the reverse von K\'{a}rm\'{a}n vortex street that commonly occurs at higher Strouhal numbers \citep{anderson1998oscillating,godoy2008transitions,godoy2009model,zhong2024predicting}, thus promoting a more coherent wake structures\citep{marais2012stabilizing,shah2024controlling}.Similar benefits have been reported for rigid foils equipped with flexible trailing edges, with numerous studies demonstrating improvements in thrust and propulsive efficiency depending on tail flexibility and operating conditions \citep{shinde2014flexibility,shinde_physics_2018,heathcote_flexible_2004,heathcote_flexible_2007,cleaver2014thrust,david2017thrust}. \\

The role of non-uniform flexibility distribution along the chord is an important but underexplored design parameter in engineered propulsion systems. In a study directly relevant to the present work, Zeyghami \citep{zeyghami2019effect} modeled chordwise non-uniform flexibility in pitching propulsors using torsional spring hinges and demonstrated that shifting the flexure location toward the trailing edge, increasing the flexion ratio, enhances propulsive efficiency but compromises thrust production, and that combining two flexible hinges can yield simultaneous gains in both metrics. While this study addressed pitching foils using a numerical model, the analogous problem for heaving foils under quiescent conditions, directly relevant to wave-assisted propulsion, remains unaddressed. Similarly, You \citep{you2019optimization} demonstrated through coupled CFD-CSD optimization that material compliance and pitching phase angle jointly govern propulsive efficiency, reinforcing the importance of simultaneously considering structural and kinematic parameters in propulsor design. The wake dynamics associated with these performance changes are also of engineering significance: Siala \citep{siala_characterization_2016} showed through PIV measurements that trailing-edge flexibility fundamentally reorganizes the vortex dynamics in the near wake of an oscillating foil, with leading, and trailing-edge compliance producing qualitatively distinct wake structures that correlate with efficiency. These findings collectively establish that flexibility placement, rather than flexibility magnitude alone, is a key design parameter in engineered propulsion systems.\\


In the present study, we focus specifically on the performance characteristics of a flexible trailing edge appended to an otherwise rigid foil. The influence of flexible tails has been previously examined for both pitching \citep{shinde2014flexibility,shinde_physics_2018,david2017thrust} and heaving foils \citep{heathcote_flexible_2004,heathcote_flexible_2007,cleaver2014thrust} under freestream and quiescent conditions. Shinde et al.\citep{shinde2014flexibility,shinde_physics_2018} investigated the effects of a flexible tail on a pitching foil in quiescent flow by fixing the flexural rigidity, $EI$ (where $E$ is Young’s modulus and $I$ is the second moment of area), while varying the pitching amplitude and frequency. Depending on the operating conditions, the flexible tail was shown to suppress the wake asymmetry commonly observed in quiescent environments \citep{shinde2013jet} and to enhance jet longevity \citep{shinde2014flexibility}. Although direct force measurements were not reported, the jet longevity was hypothesized to correlate with enhanced thrust production. David et al.\citep{david2017thrust} reported complementary force and flow measurements for a pitching foil with flexible tail under free-flow conditions. In that study, both the flexural rigidity and tail length were varied, enabling exploration over a wide range of the non-dimensional flexural rigidity parameter, $R^*$, defined as
\begin{equation}\label{eq:Rstar}
R^* = \frac{EI}{0.5\rho U_\infty^2 c_f^3},
\end{equation}
where $\rho$ is the fluid density, $U_\infty$ is the freestream velocity, and $c_f$ is the length of the flexible tail. The results demonstrated that the maximum propulsive efficiency occurred at $R^* \approx 0.01$, while the maximum thrust production was achieved at $R^* \approx 8$.\\

The role of flexible tails has also been examined for purely heaving foils in both freestream \citep{heathcote_flexible_2007} and quiescent conditions \citep{heathcote_flexible_2004}. These studies reported that thrust production was maximized at moderate tail flexibility, while higher flexibility generally resulted in increased efficiency. Similarly, Cleaver \citep{cleaver2014thrust} investigated the influence of flexible-tail length for a heaving foil in freestream flow and found that moderate trailing-edge deflections produced higher thrust than cases with larger deflections. \Revone{The prospect of using flexibility distribution as a practical design lever, without altering total chord, flexural rigidity, or actuation architecture, is particularly attractive for wave-driven USV propulsors, where structural simplicity and endurance are primary engineering objectives \citep{bianchi2022numerical}.} While these studies provide important insights into the role of flexible tails in heaving-foil propulsion, a more comprehensive exploration of the combined effects of flexibility, flexure location, and operating conditions is necessary. Such systematic characterization is essential for identifying optimal flexibility parameters and informing the effective integration of flexible foils into wave-assisted propulsion systems.\\

As an initial step toward this objective, the present study examines the effects of flexible-tail length on a purely heaving foil operating at quiescent flow conditions. Although Heathcote et al.\citep{heathcote_flexible_2004} investigated flexible tails for heaving foils in quiescent flow, that study varied only the flexural rigidity while holding the tail length fixed. Previous work has shown  that tail length is also a key parameter that influences wake dynamics and propulsive performance \citep{cleaver2014thrust,david2017thrust}. As such, we investigate the effects of length of the flexible tail on heaving foil in quiescent by varying the flexure location while keeping both the airfoil chord and the overall chord length constant. Specifically, we aim to address the following research questions:
\begin{enumerate}
    \item How does the flexible tail length influence the performance of a heaving foil in quiescent conditions?
    \item How do wake characteristics correlate with the observed performance behavior in quiescent conditions?
\end{enumerate}
To answer these questions, experiments are conducted at a range of heave amplitudes and frequencies, and simultaneous force and flow-field measurements are performed to establish direct correlations between performance metrics and wake characteristics. \Revone{The results provide physics-based design criteria for selecting flexibility distribution in wave-driven propulsor development, with direct relevance to the engineering of efficient, low-emission unmanned marine vehicles.} Details of the experimental setup and operating conditions are provided in \S\ref{sec:methods}. The results are presented in \S\ref{sec:results}, followed by a discussion of their implications and broader relevance in \S\ref{sec:discussion}. The main conclusions are summarized in \S\ref{sec:conclusion}.

\section{Experimental Methods and Operating Conditions}
\label{sec:methods}
\begin{figure}[h!]
  \centerline{\includegraphics[width=0.8\columnwidth]{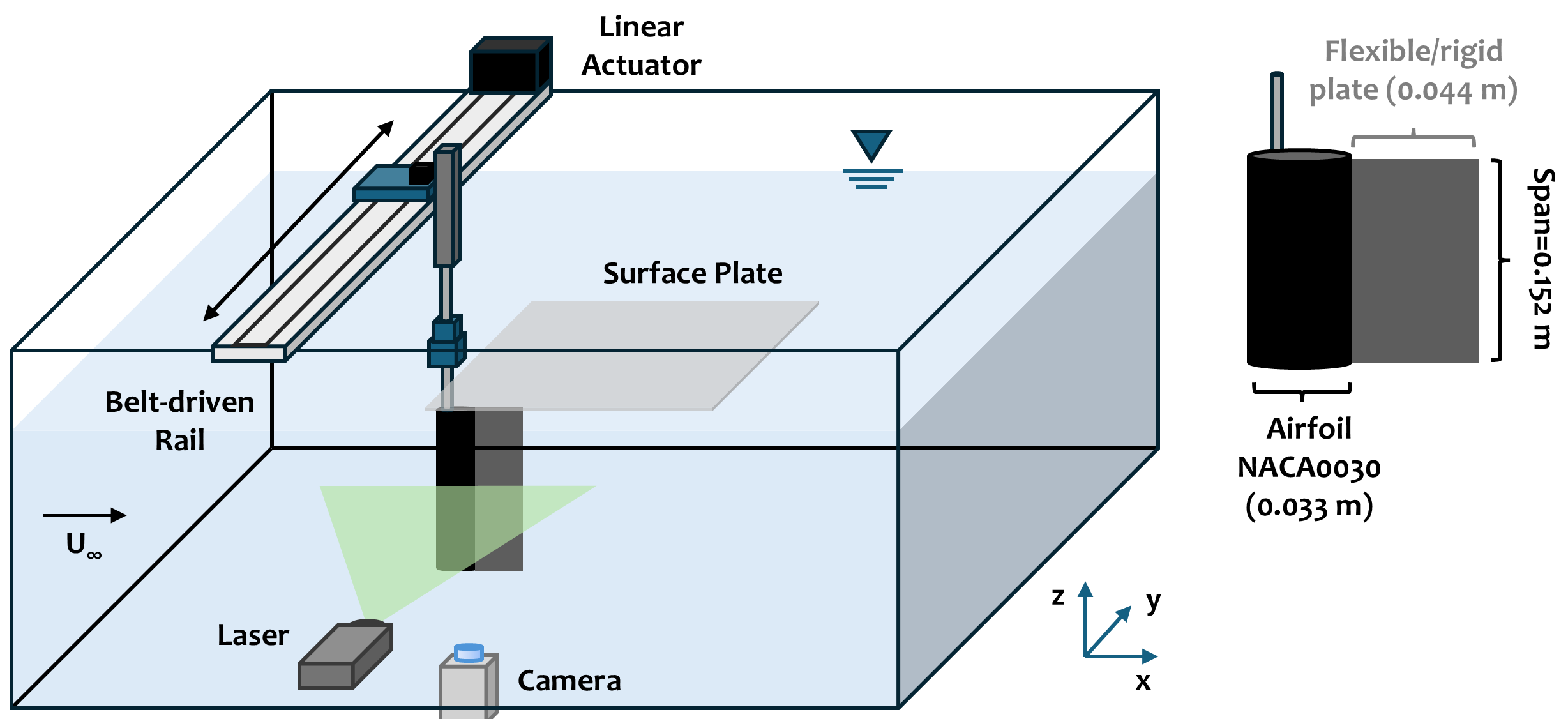}}
  \caption{A brief schematic of the experimental setup used in the current study.}
\label{fig:facility}
\end{figure}

\vspace{-0.1in}
The schematic representation of the experimental facility is shown in Fig.~\ref{fig:facility}. The experiments were carried out in a water channel with a $0.6~\mathrm{m}$ wide, $0.8~\mathrm{m}$ high, and $3~\mathrm{m}$ long test section. All experiments were carried out under quiescent conditions with zero freestream. Following~\cite{raut2024hydrodynamic}, the wave-induced motion was simulated using sinusoidal excitation, which was provided in the current study using a linear traverse driven by a NEMA~23 stepper motor. The motion conformity was verified using a rotary optical incremental encoder (E5 US Digital) mounted on the transmission shaft. The transmission shaft was instrumented with  ATI Mini40, a six-axis force sensor with a resolution of  $0.01~\mathrm{N}$ in force and $0.00025~\mathrm{Nm}$ in moment to characterize the foil performance. Custom in-house adapters were used to mount the foils to the transducer. The foil consisted of a NACA0030 airfoil of chord length $c_r=0.033~\mathrm{m}$ with an attached thin tail of length $0.044~\mathrm{m}$, resulting in an overall chord length of $c_t=0.077~\mathrm{m}$. The overall chord $c_t$ was used as the characteristic length scale for normalization, consistent with previous studies \citep{heathcote_flexible_2004,david2017thrust}. \\

\begin{table}[]\centering
\caption{Summary of the operating conditions}
\label{tab:operatingConditions}
\begin{tabular}{cccll}
\cline{1-3}
Flexible tail length (m) & Heave frequency (Hz)       & Heave amplitude-chord ratio &  &  \\ \cline{1-3}
0 (Rigid)                & \multirow{3}{*}{0.8, 1.25} & \multirow{3}{*}{0.13, 0.22} &  &  \\
0.022 (Mid-flexible)     &                            &                             &  &  \\
0.044 (Fully-flexible)    &                            &                             &  &  \\ \cline{1-3}
\end{tabular}
\end{table}
The foil span was $s=0.152~\mathrm{m}$, corresponding to aspect ratios of $AR_{c_r}=4.6$ and $AR_{c_t}=2$ based on the airfoil chord and overall chord, respectively. A surface plate was positioned at the free surface, with a clearance of $\leq2~\mathrm{mm}$ from one end of the foil to suppress surface-wave effects. This plate imposed an effective symmetry condition, doubling the effective aspect ratio \citep{zhu2023flow}. Consequently, three-dimensional effects are expected to be minimal near the mid-span, where particle image velocimetry (PIV) measurements were acquired.

For the control case (hereafter referred to as the ``rigid'' case), the NACA0030 airfoil was fitted with a rigid flat-plate tail of thickness $t = 0.79~\mathrm{mm}$ ($1/32 \ \mathrm{in}$). The airfoil was fabricated using Vero rigid material on a Stratasys J850 3D printer with a resolution of 25 $\mu$m. To investigate the effect of structural flexibility, the rigid tail was replaced with a high-strength neoprene sheet of identical thickness. The location of the flexure was varied between $100\%$ and $50\%$ of the tail length, thereby altering the extent of the flexible section. In the $100\%$ case, a neoprene sheet with chord length $c_f = 0.044~\mathrm{m}$ was directly attached to the trailing edge of the NACA0030 airfoil. This configuration is referred to as the ``fully flexible'' case throughout this manuscript. In the $50\%$ case, a rigid flat plate of length $0.022~\mathrm{m}$ was first attached to the trailing edge of the airfoil, followed by a neoprene sheet with chord length $c_f = 0.022~\mathrm{m}$. This configuration is referred to as the ``mid flexible'' case.\\

The neoprene sheet utilized in these experiments featured a Shore A hardness of 50. The Elastic Modulus ($E$) of the neoprene was estimated to be $E = 3.371 \pm 0.081~\mathrm{MPa}$. The second moment of area of the flap cross-section was calculated as $I = \frac{1}{12} bt_f^3$, where $b$ is the span of the foil and $t_f$ is the flap thickness, resulting in $I = 6.35 \times 10^{-12}~\mathrm{m^4}$. As a result, the flexural rigidity of the tail was estimated to be $EI = (2.14 \pm 0.05) \times 10^{-5}~\mathrm{Nm^2}$.\\

The heave amplitude was varied as $h/c_t = 0.13$ and $0.22$, and the heave frequency was set to $f = 0.8$ and $1.25~\mathrm{Hz}$. The resulting Reynolds numbers, based on the peak heave velocity $v_h = 2\pi h f$ and the overall chord length $c_t$, were $Re = 3850$, $6006$, $6545$, and $10010$. The summary of the operating conditions is shown in Table~\ref{tab:operatingConditions}. Each test was repeated four times to ensure repeatability. The confidence intervals in the figures represent the standard deviation ($\sigma$) of the mean across the four trials. Under these operating conditions, the motion of the tail was quantified using optical tracking. To obtain sufficient contrast against the surroundings, reflective paint was applied at the trailing-edge of the tail. Using a digital camera, the motion was recorded at a frame rate of $30~\mathrm{Hz}$. The recorded image sequences were then post-processed using a custom MATLAB algorithm based on intensity thresholding to extract the temporal evolution of the trailing-edge displacement.
Nomenclature of all the variables and parameters used in the current study are tabulated in table~\ref{tab:nomenclature}. 

\begin{table}[]
\centering
\caption{Nomenclature of variables and parameters used in this study.}
\label{tab:nomenclature}
\begin{tabular}{cl@{\hspace{2cm}}cl}

\multicolumn{2}{c}{Symbol \& Description} & \multicolumn{2}{c}{Symbol \& Description} \\ \cline{1-2} \cline{3-4}

$c_r$           & NACA0030 foil chord                & $t$                 & Thickness of the tail \\
$c_f$           & Flexible tail length                    & $v_h$               & Peak heave velocity \\
$c_t$           & Total foil length              & $\nu$               & Fluid kinematic viscosity \\
$f$             & Heave frequency                      & $C_T$               & Thrust coefficient \\
$h^*$           & Heave-to-chord ratio           & $\epsilon$          & Thrust-to-power ratio \\
$h$             & Heaving amplitude              & $u_{max}$           & Peak streamwise jet velocity \\
$R^*$           & Dimensionless flexural rigidity      & $\omega$            & Vorticity \\
$E$             & Modulus of elasticity          & $\delta_{TE}$       & TE deflection \\
$I$             & Second moment of area          & $\delta_{h,TE}$     & Cross-stream TE deflection \\
$\rho$          & Fluid density                        & $\delta_{x,TE}$     & Streamwise TE deflection \\
$U_\infty$      & Free-stream velocity           & $\delta_{h,TE_{max}}$ & Maximum cross-stream TE deflection \\
$s$             & Span of foil              & $\phi^\circ$        & Phase lag between pitch \& heave\\
$AR_{c_r}$      & Aspect ratio of NACA foil      & $a_{TE}$            & Maximum TE displacement \\
$AR_{c_t}$      & Aspect ratio of entire foil    & $V_{TE_{max}}$      & Maximum TE velocity \\

\cline{1-2} \cline{3-4}
\end{tabular}

\end{table}

\subsection{Flow Measurements}
\label{subsec:flowfield}
The flow field was captured using 2D-2C Particle Image Velocimetry (PIV) in the $x-y$ plane, passing through the foil's midspan. The seeding was achieved using $55\mathrm{\mu m}$ polyamide particles with a specific gravity of $1.01$ g/cm$^3$. \Revone{The resulting Stokes number ($St$) for the current flow conditions was estimated to be $0.02$, which is well below the recommended value of $St\leq0.05$~\cite{samimy1991motion}}. An Evergreen 200 dual-pulsed laser with a pulse energy of 200 mJ was used to illuminate the particles. A Photron NOVA R3-4K camera with a resolution of  $4096\times2304$ pixels$^2$ was used to capture the PIV images. The camera was equipped with a 60 mm Nikon lens to provide a field of view of $2.7c_t\times1.5c_t$. A total of 1080 image pairs were captured over 72 seconds at a sampling rate of 15 Hz for all cases. The time delay between pulses ($\delta t$) was optimized for each case to ensure accurate velocity measurements. The images were captured and processed using DaVis 11 (LaVision) software. The vector field was obtained by processing the images using a multi-grid, multi-pass, cross-correlation technique with an initial and final interrogation window of size $96\times96$ pixels$^2$ and $64\times64$ pixels$^2$, respectively, with an overlap of $50\%$. The resulting vector field had a spatial resolution of $1.64~\mathrm{mm}$~($0.021c_t$). The uncertainties in the velocity measurements were quantified using the correlation statistics method~\citep{Wieneke_2015}, where the uncertainty is proportional to the residual positional disparity between the matched correlation peaks~\citep{Sciacchitano_2019}. The resulting uncertainty in the velocity across the flowfield for all cases was $\le1\%$ of the average velocity in the flow field.\\

The velocity vector fields were post-processed in MATLAB to obtain both time-averaged and phase-averaged quantities. For the time-averaged analysis, the vector fields were averaged over cycles 20–50 for each experimental run and subsequently averaged across four independent repetitions. The standard deviation reported for the time-averaged flow fields corresponds to the variability across these four runs. Averaging was restricted to cycles beyond the $20^{th}$ cycle, as flow convergence was consistently observed only after this point. Phase-averaged flow fields were computed using encoder data synchronized with the PIV measurements to define the phase reference. Similar to the time-averaged analysis, phase averaging was performed over cycles 20–50 to ensure statistical convergence.

\section{Results}\label{sec:results}
\subsection{Validation}
\begin{figure}
     \centering
     \begin{subfigure}{0.5\textwidth}
         \centering
         \includegraphics[width=0.9\textwidth]{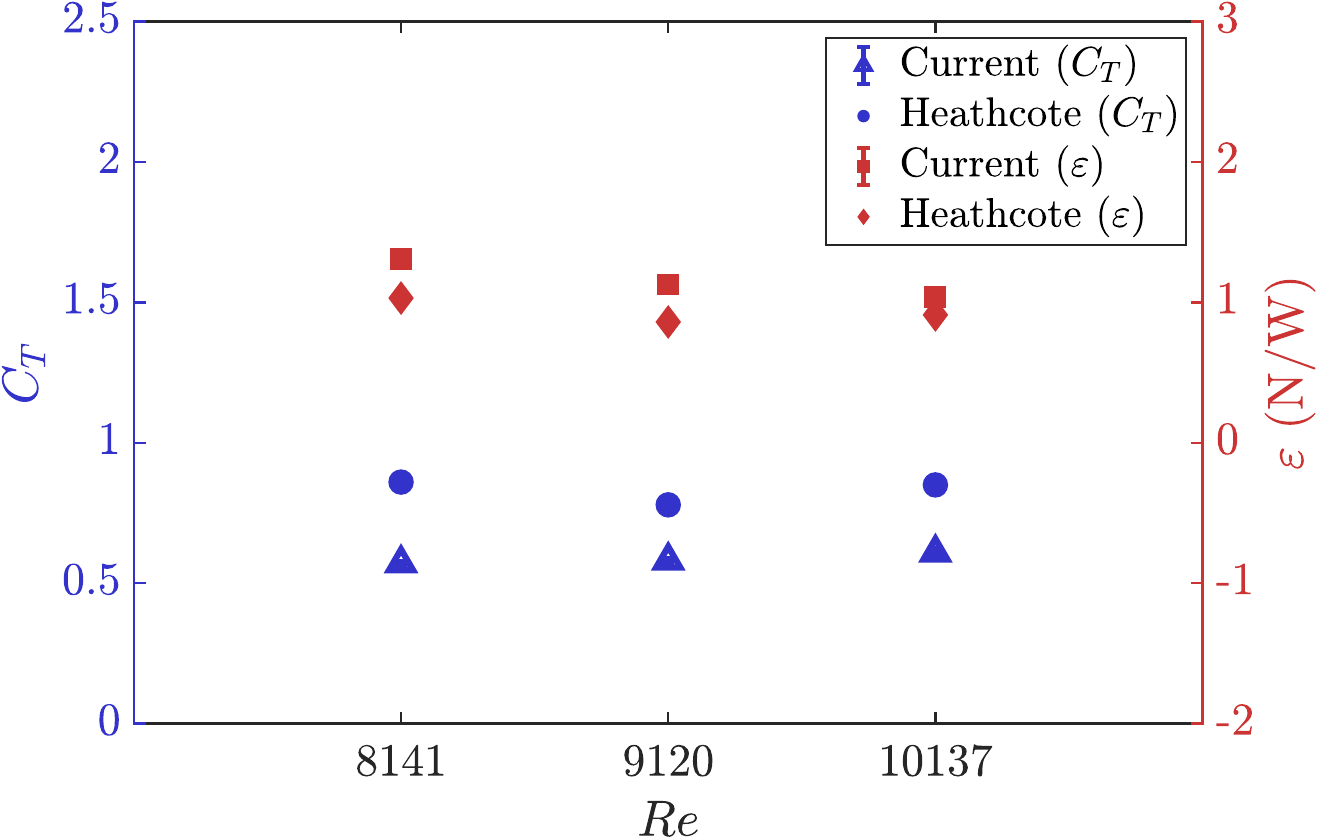}
         \caption{}
         \label{subfig:validationHeathcote}
     \end{subfigure}%
     \hfill
     \begin{subfigure}{0.5\textwidth}
         \centering
         \includegraphics[width=0.9\textwidth]{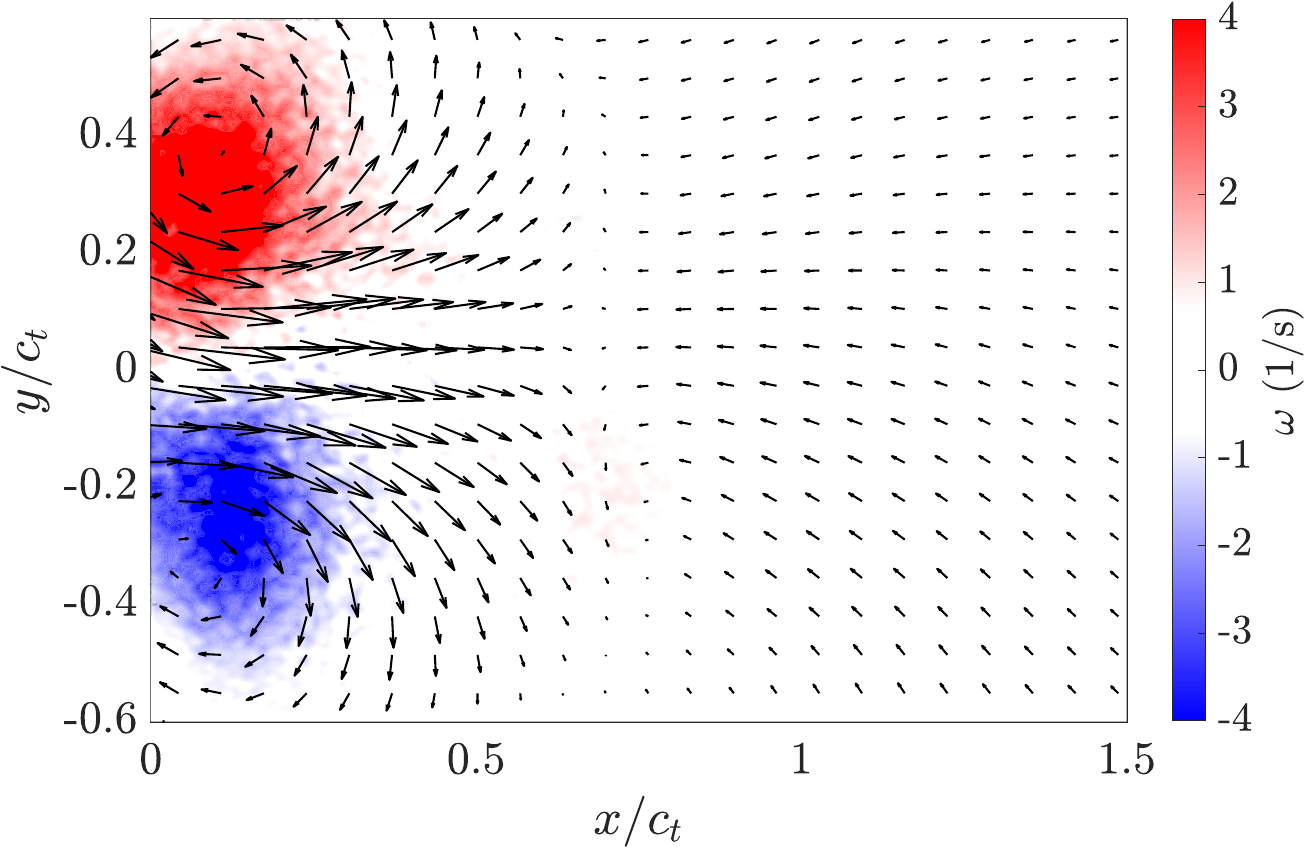}
         \caption{}
         \label{subfig:timeAveragedRigidValidation}
     \end{subfigure}%
        \caption{(a) Comparison of thrust coefficient ($C_T$) and thrust-to-power input ratio ($\epsilon$) between the current measurements and that by Heathcote et al.~\cite{heathcote_flexible_2004}. (b) Time-averaged vorticity field overlaid with time-averaged streamlines for the rigid case at heave frequency and amplitude of $1.25~\mathrm{Hz}$ and $h^*=0.22$, respectively.}
        \label{fig:validationHeathcote}
\end{figure}

The performance of the rigid control case was compared against the measurements of Heathcote and Gursul~\cite{heathcote_flexible_2004} to benchmark the present experimental facility. Their study provides the closest relevant dataset, using a thin flat plate of length $60~\mathrm{mm}$ attached to a teardrop-shaped airfoil with chord length $30~\mathrm{mm}$ and thickness $10~\mathrm{mm}$ in quiescent flow. To closely replicate their operating conditions, particularly the Reynolds number based on the total chord, $Re = f c_t^2/\nu$, the heave ratio was set to $h/c_t = 0.194$, and the heave frequencies were chosen as $1.37$, $1.54$, and $1.7~\mathrm{Hz}$. The thrust coefficient is defined as $C_T = \overline{T}/{\frac{1}{2}\rho v_p^2 c s}$,
where $\overline{T}$ is the time-averaged thrust force. Following Frampton et al.~\cite{frampton2001passive}, the thrust-to-power-input ratio, $\epsilon$, is defined as the ratio of average thrust to input power. Note here that the conventional definition of propulsive efficiency can not be used due to absence of freestream velocity. Thus, following ~\cite{heathcote_flexible_2004, frampton2001passive}, we define propulsive efficiency as:
$$\epsilon = \overline{T} (N) / F v ,$$ where $\overline T$ is the time-averaged thrust force in Newtons and $F v$ represents power consumed in prescribing heave motion. This quantity has units of newton/watt. Both quantities are plotted in Fig.~\ref{fig:validationHeathcote} and compared directly with the data reported by Heathcote and Gursul~\cite{heathcote_flexible_2004}. The results show good overall agreement in trend, with only modest differences in the thrust magnitude.
The observed discrepancies can be attributed to differences in geometry and boundary conditions between the two studies. In the present work, the flap-to-chord ratio is smaller ($c_f/c_t = 0.57$) compared to $0.67$ in Heathcote and Gursul~\cite{heathcote_flexible_2004}. Furthermore, their experiments employed end plates to approximate an infinite-span configuration, whereas the present measurements utilized an end plate only at the top to mitigate free-surface effects, with the lower end remaining free. The combined influence of flap geometry and finite-span three-dimensional effects is therefore expected to contribute to the observed differences.\\

To characterize the wake, the time-averaged flow field at the midspan of the rigid heaving foil was measured using particle image velocimetry. Figure~\ref{subfig:timeAveragedRigidValidation} shows the time-averaged vorticity field overlaid with time-averaged streamlines. The wake is characterized by a closely spaced dipole comprising two counter-rotating vortices that decay within approximately one chord length downstream. This near-wake organization and rapid jet decay are consistent with previous observations for rigid pitching and heaving foils in quiescent flow \citep{Silwal2026, heathcote2007jet, heathcote_flexible_2004}. In particular, our previous measurements on a rigid pitching foil showed that destructive interactions between closely spaced counter-rotating vortices limit jet persistence under zero-freestream conditions \cite{Silwal2026}. This behavior contrasts with finite-span flapping foils, which typically generate vortex-loop and vortex-ring wake topologies due to pronounced spanwise flow and tip-vortex dynamics \citep{dong2006wake, buchholz2008wake}. Although three-dimensional effects are expected to be present for a finite-span foil with an effective aspect ratio of 4, these are primarily confined to the tip region. The midspan plane, where the PIV measurements are performed, presumably remains largely unaffected by tip-induced three-dimensionality and exhibits the dominant dipolar wake structure characteristic of two-dimensional rigid flapping foils in quiescent flows. Accordingly, while three-dimensional effects are present, midspan two-dimensional PIV captures the primary flow physics and provides a consistent basis for quantifying relative differences across the chordwise flexibility cases examined in this study. \\

It is also noted that rigid flapping foils commonly exhibit jet deflection \citep{shinde2013jet, heathcote2007jet}, which is not observed here. This discrepancy is likely due to three-dimensional effects, as jet deflection is predominantly a two-dimensional phenomenon \cite{calderon2014}, whereas spanwise flow and tip-induced mixing in finite-span configurations tend to suppress such deflections. To further assess the influence of three-dimensionality on performance, additional experiments were conducted with end plates to approximate infinite-span conditions; details are provided in the Appendix. The results show only minor differences in thrust and efficiency between finite- and infinite-span configurations. 
While three-dimensionality alters the wake structure, its effect on the integrated performance metrics appears limited. Moreover, even under infinite-span conditions, the measured thrust remains slightly lower than that reported by Heathcote (see Appendix), likely due to differences in tail length. Accordingly, all subsequent comparisons among flexibility cases are performed under finite-span conditions.

\subsection{Performance characteristics}
\begin{figure}
     \centering
     \begin{subfigure}{0.5\textwidth}
         \centering
         \includegraphics[width=0.95\textwidth]{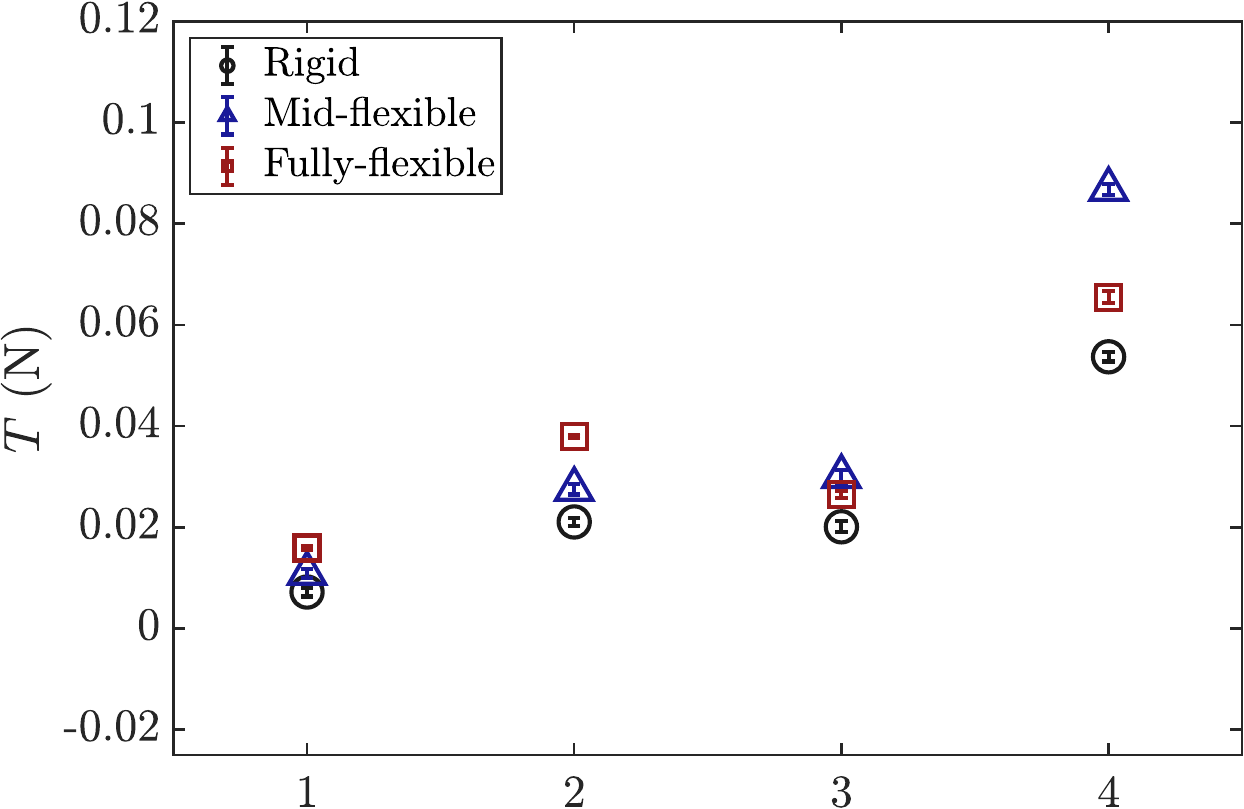}
         \caption{}
         \label{subfig:thrustComparison}
     \end{subfigure}%
     \hfill
     \begin{subfigure}{0.5\textwidth}
         \centering
         \includegraphics[width=0.9\textwidth]{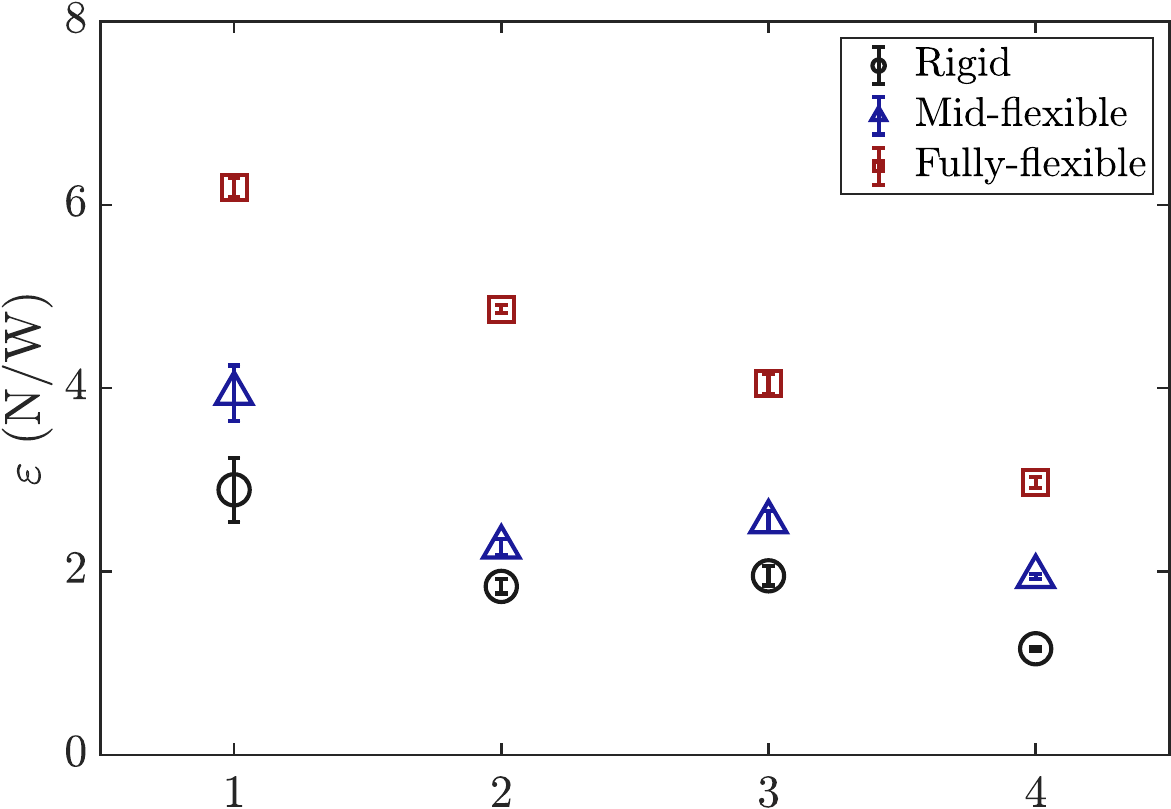}
         \caption{}
         \label{subfig:powerComparison}
     \end{subfigure}%

        \caption{Comparison of (a) thrust, and (b) thrust-to-power input ratio ($\epsilon$) between rigid, mid-flexible and fully-flexible cases. The labels in the x-axis represents different operating conditions, where 1, 2, 3, and 4 corresponds to ($h^*=0.13,f = 0.8~\mathrm{Hz}$), ($h^*=0.22,f = 0.8~\mathrm{Hz}$), ($h^*=0.13,f = 1.25~\mathrm{Hz}$), and ($h^*=0.22,f = 1.25~\mathrm{Hz}$), respectively.}
        \label{fig:CTandEfficiency}
\end{figure}

The performance characteristics of the three configurations across the four operating conditions are summarized in Fig.~\ref{fig:CTandEfficiency}. \Cref{subfig:thrustComparison,subfig:powerComparison} show the dimensional thrust ($T (N)$) and thrust-to-power input ratio ($\epsilon (N/W)$), respectively. Conditions 1 and 2 correspond to a heaving frequency of $0.8~\mathrm{Hz}$ with amplitudes $h^* = 0.13$ and $0.22$, while conditions 3 and 4 correspond to a frequency of $1.25~\mathrm{Hz}$ with the same amplitudes. As expected, both thrust and power increase with increasing heave frequency and amplitude.\\

Among the three tail-rigidity configurations, the rigid tail consistently yields the lowest thrust and requires the highest power input across all operating conditions. Between the two flexible cases, the fully-flexible tail produces marginally higher thrust at the lower-frequency conditions (cases 1 and 2), whereas the mid-flexible tail generates higher thrust at the higher-frequency conditions (cases 3 and 4). Notably, the thrust enhancement for the mid-flexible case is most pronounced at condition 4. This dependence of optimal rigidity on operating conditions is consistent with the observations of Heathcote \cite{heathcote_flexible_2004}. In contrast, the fully-flexible tail exhibits a higher thrust-to-power ratio across all conditions, indicating superior propulsive efficiency relative to the other configurations. This trend is again consistent with the findings of Heathcote \cite{heathcote_flexible_2004}, who reported enhanced efficiency for foils with greater flexibility, and with the results of David \cite{david2017thrust}, who demonstrated that optimal thrust and efficiency depend jointly on structural rigidity and kinematics. To elucidate the mechanisms underlying these performance trends, the time-averaged and phase-averaged flowfields, together with the measured tail kinematics, are examined in the following sections.

\begin{figure}[h!]
  \centerline{\includegraphics[width=1\columnwidth]{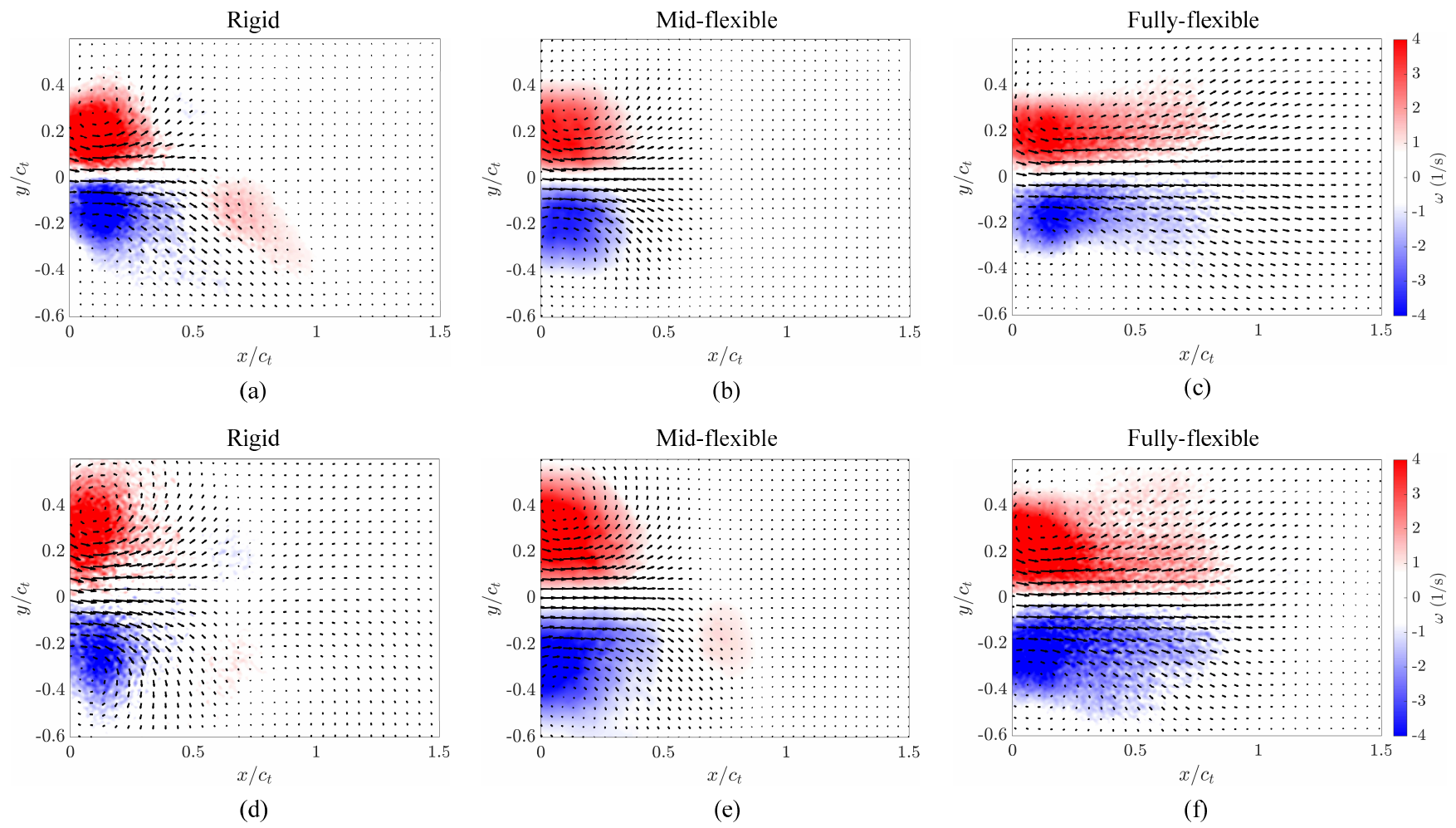}}
  \caption{Time-averaged vorticity field overlaid with time-averaged streamlines for rigid, mid-flexible and fully-flexible cases at heave frequency and amplitude of (a-c) $1.25~\mathrm{Hz}$ and $h^*=0.13$, and (d-e) $1.25~\mathrm{Hz}$ and $h^*=0.22$. }
\label{fig:timeAveragedFlowfield}
\end{figure}

\subsection{Flowfield characteristics}
\begin{figure}
     \centering
     \begin{subfigure}{0.5\textwidth}
         \includegraphics[width=0.9\textwidth]{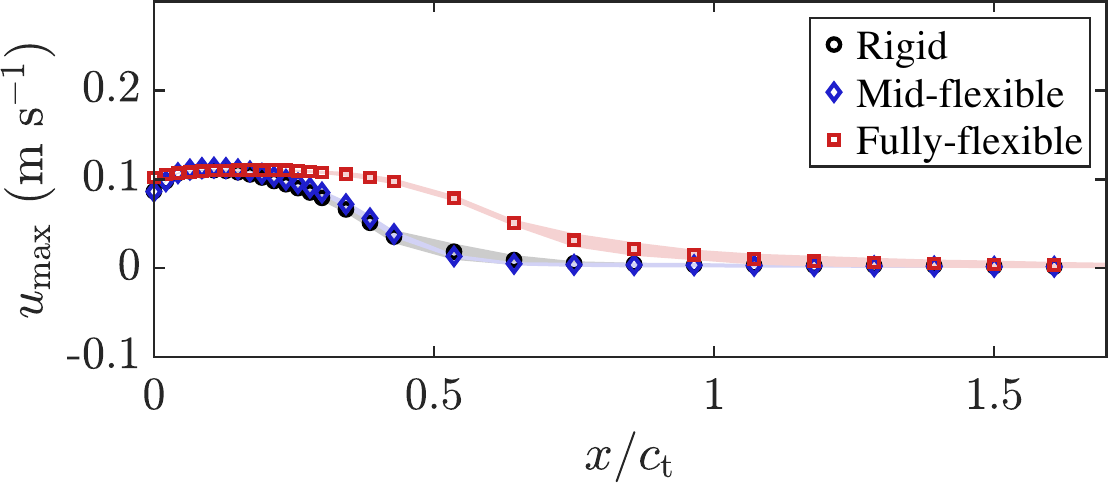}
         \caption{}
         \label{subfig:08Hz10mmvelocityMax}
     \end{subfigure}%
     \hfill
     \begin{subfigure}{0.5\textwidth}
         \centering
         \includegraphics[width=0.9\textwidth]{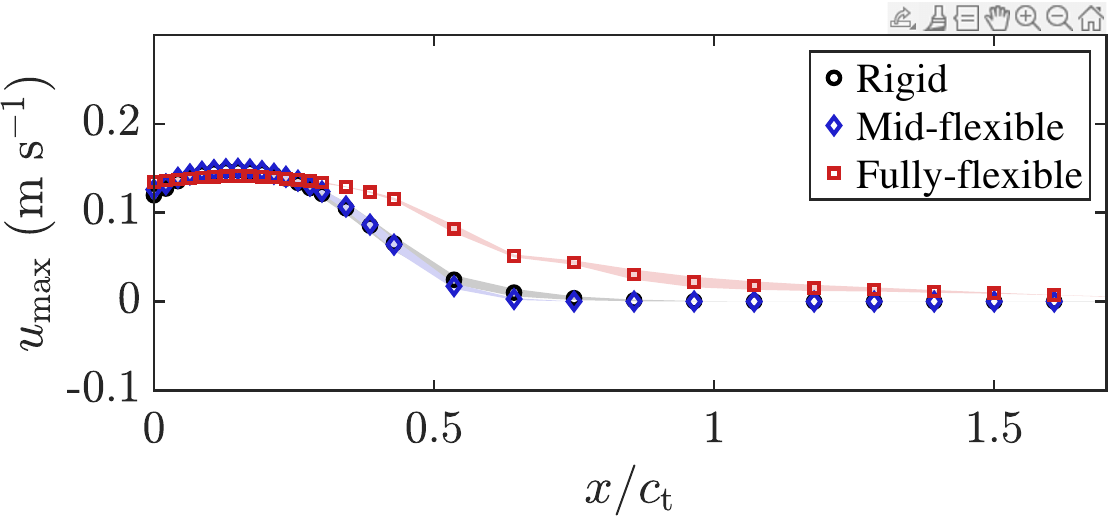}
         \caption{}
         \label{subfig:08Hz17mmvelocityMax}
     \end{subfigure}%

     \centering
     \begin{subfigure}{0.5\textwidth}
         \includegraphics[width=0.9\textwidth]{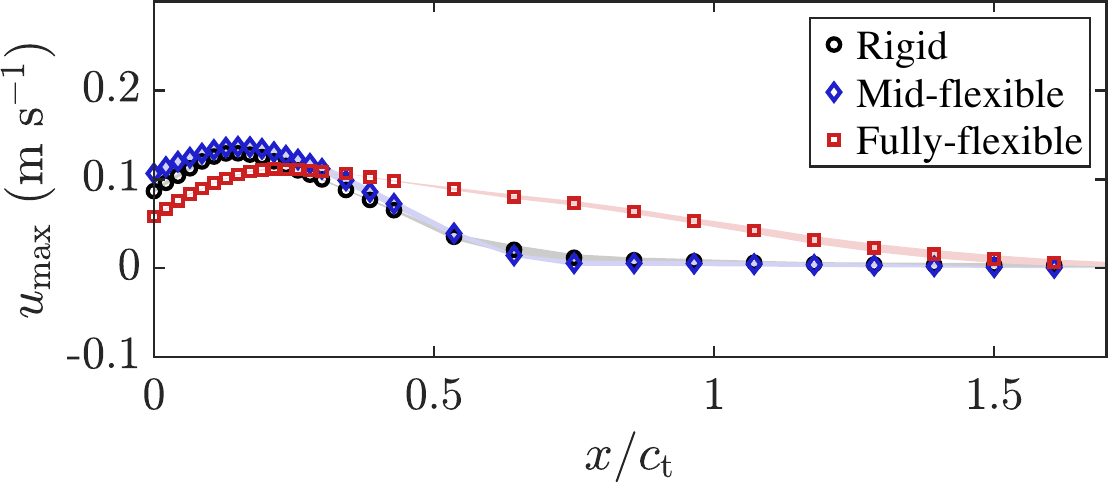}
         \caption{}
         \label{subfig:125Hz10mmvelocityMax}
     \end{subfigure}%
     \hfill
     \begin{subfigure}{0.5\textwidth}
         \centering
         \includegraphics[width=0.9\textwidth]{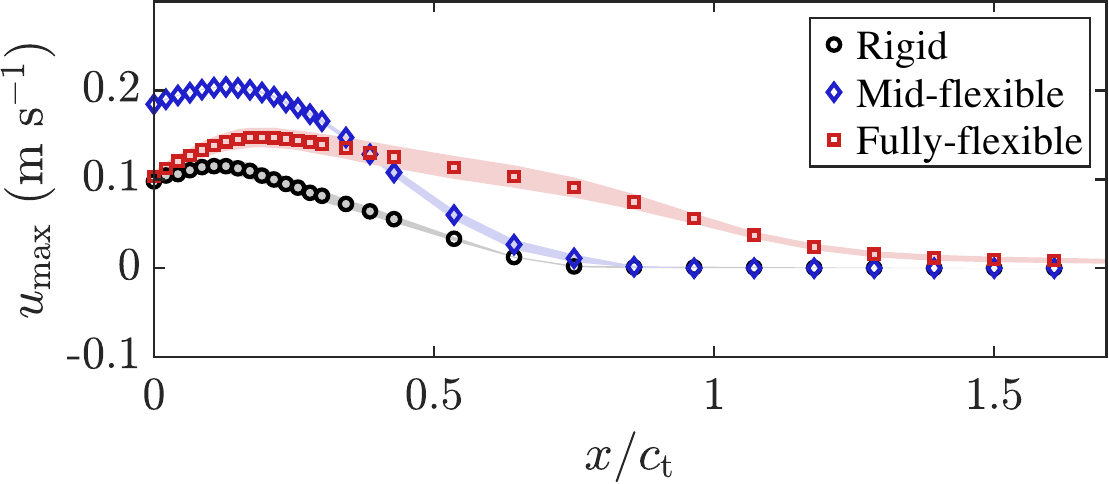}
         \caption{}
         \label{subfig:125Hz17mmvelocityMax}
     \end{subfigure}%
        \caption{Comparison of the peak streamwise jet velocity ($u_{max}$) between rigid, mid-flexible and fully-flexible cases at (a) $0.8~\mathrm{Hz},~h^*=0.13$, (b) $0.8~\mathrm{Hz},~h^*=0.22$, (c) $1.25~\mathrm{Hz},~h^*=0.13$, and (d) $1.25~\mathrm{Hz},~h^*=0.22$. The shaded region represents the standard deviation across four repetitions.}
        \label{fig:meanAndMaxVelocities}
\end{figure}

The time-averaged vorticity fields for the three foils at conditions 3 ($h^* = 0.13$, $f$ = 1.25 Hz) and 4 ($h^* = 0.22$, $f$ = 1.25 Hz) are shown in Fig.~\ref{fig:timeAveragedFlowfield}. In all of the cases, the wake is characterized by a dipole structure, with several notable features. First, the size of the vortex and the spanwise spacing between the two vortices in the dipole increase with heave amplitude. This is correlated with higher thrust values for $h^* = 0.22$ compared to $h^* = 0.13$ for all three foils, as shown in figure~\ref{fig:CTandEfficiency}a. Second, the jet is sustained for longer in the fully-flexible case for both the heave amplitudes. This is correlated with higher propulsion efficiency for the fully-flexible foil, as shown in figure~\ref{fig:CTandEfficiency}b.
To quantify this effect, the peak streamwise jet velocity ($u_{\max}$) is extracted from the time-averaged flowfields as a metric of jet strength, with larger values indicating stronger momentum flux in the wake. The variation of $u_{\max}$ with operating conditions and flexibility is shown in Fig.~\ref{fig:meanAndMaxVelocities}.\\

\begin{figure}[h!]
  \centerline{\includegraphics[width=0.85\columnwidth, trim={0cm 3cm 0cm 2cm}]{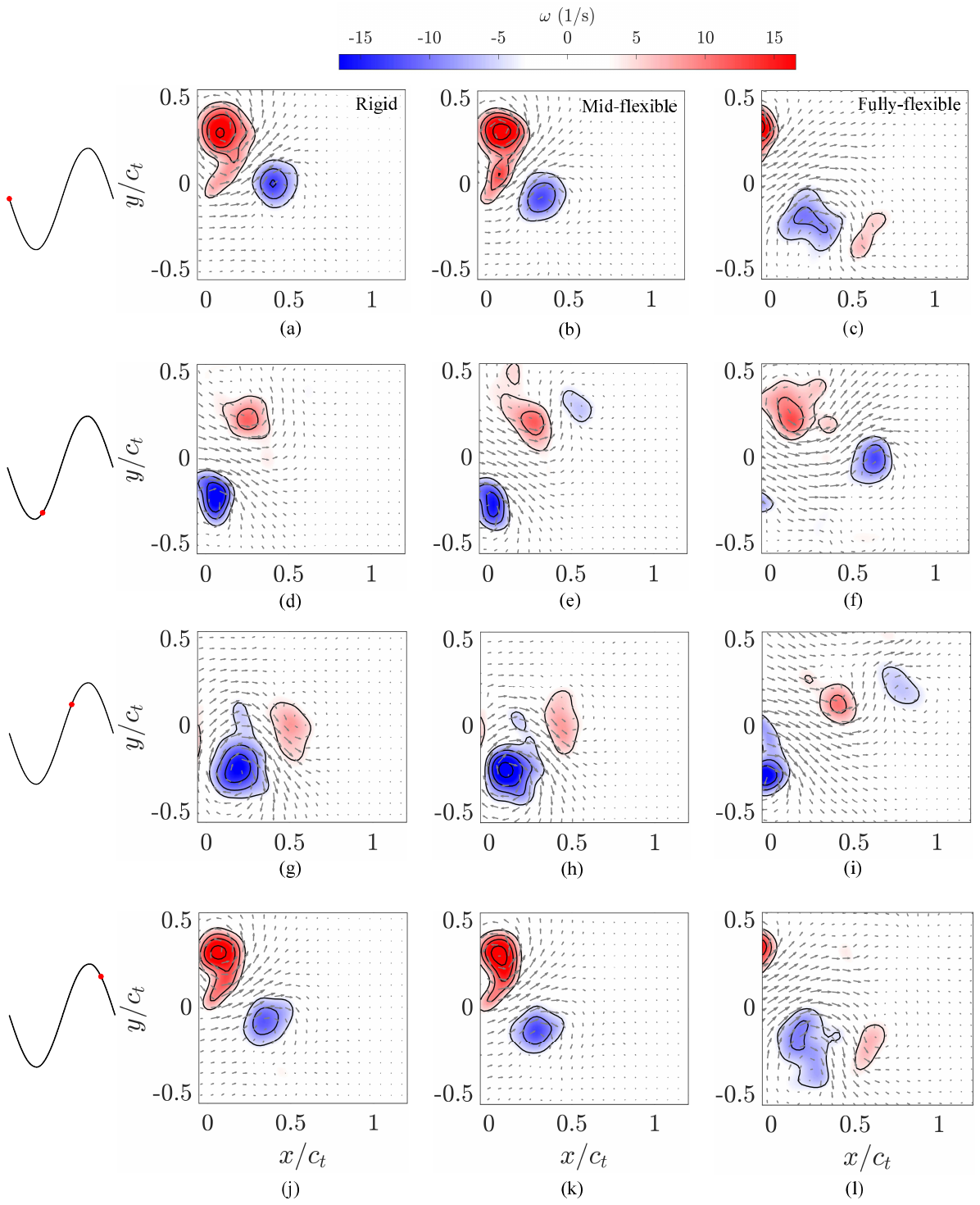}}
  \caption{Comparison of the phase-averaged vorticity fields between rigid, mid-flexible and fully-flexible cases at different phases of the flapping cycle. These results are for the heave frequency and amplitude of $0.8~\mathrm{Hz}$ and $h^*=0.22$, respectively.}
\label{fig:085HzPhaseAveraged}
\end{figure}

Across all operating conditions, the fully-flexible case exhibits higher jet sustainment relative to the other cases, as indicated by the slower downstream decay of peak velocities. This is presumably due to increased spacing between the shed trailing-edge vortices, which reduces the likelihood of destructive vortex–vortex interactions and allows coherent structures to persist further downstream. Increased vortex spacing associated with flexible trailing edges has been previously reported for both pitching \citep{shinde2014flexibility,david2017thrust} and heaving foils \citep{heathcote_flexible_2004}. A similar relationship between increased vortex spacing, reduced destructive vortex interactions, and prolonged jet persistence was observed in our previous study of a rigid pitching foil with surface dimples \cite{Silwal2026}. Although flexibility and surface roughness modify the flow through different physical mechanisms, both results demonstrate that the timing and spacing of trailing-edge vortex shedding govern the downstream persistence of the momentum jet. The resulting wake resembles a reverse von K\'{a}rm\'{a}n vortex street with two coherent trailing-edge vortices shed per cycle, a structure commonly associated with efficient thrust generation. The enhanced efficiency observed for the fully-flexible case is therefore consistent with the corresponding wake dynamics. Similarly, the jet decay characteristics between the rigid and mid-flexible cases were nearly identical for all the operating conditions tested, and thus, the power requirements were similar between them; higher $\epsilon$ for the mid-flexible case is due to slightly higher thrust for similar power consumption. This suggests that the vortices interact and decay faster for these cases and are less efficient compared to the fully-flexible case.\\ 

To further assess vortex longevity and its connection to propulsive efficiency, phase-averaged flowfields at different phases within a heave cycle were compared among the three rigidity cases at $f = 0.8~\mathrm{Hz}$ and $h^* = 0.22$, as shown in Fig.~\ref{fig:085HzPhaseAveraged}. For both the rigid and mid-flexible cases, the shed vortices form a closely spaced dipole configuration, characteristic of higher-Strouhal-number wakes. Such wake structures are typically associated with reduced propulsive efficiency due to strong vortex–vortex interactions and rapid wake decay. In contrast, the fully-flexible case exhibits substantially larger streamwise spacing between successive vortices, allowing them to convect farther downstream and sustain a coherent jet over a longer distance.\\

\begin{figure}
     \centering
     \begin{subfigure}{0.5\textwidth}
         \includegraphics[width=0.9\textwidth]{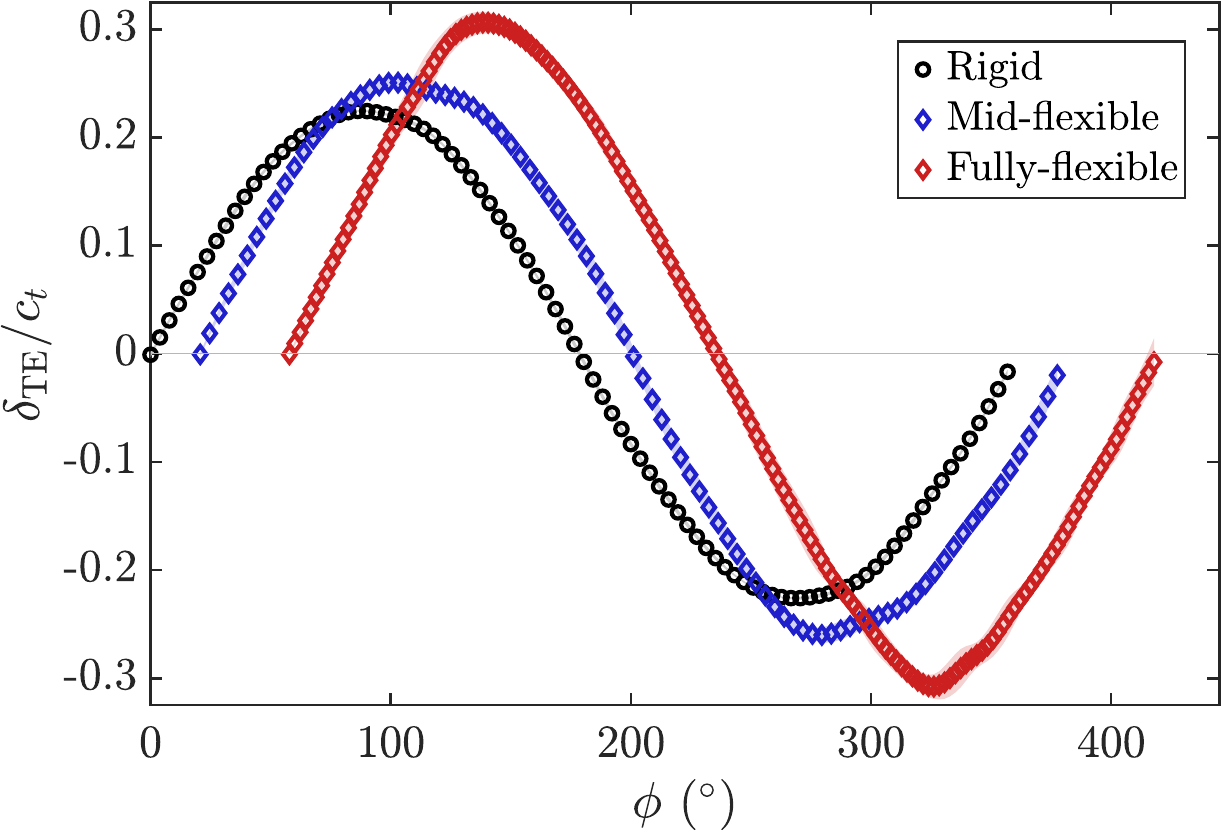}
         \caption{}
         \label{subfig:trackingHOneCondition}
     \end{subfigure}%
     \hfill
     \begin{subfigure}{0.5\textwidth}
         \centering
         \includegraphics[width=0.9\textwidth]{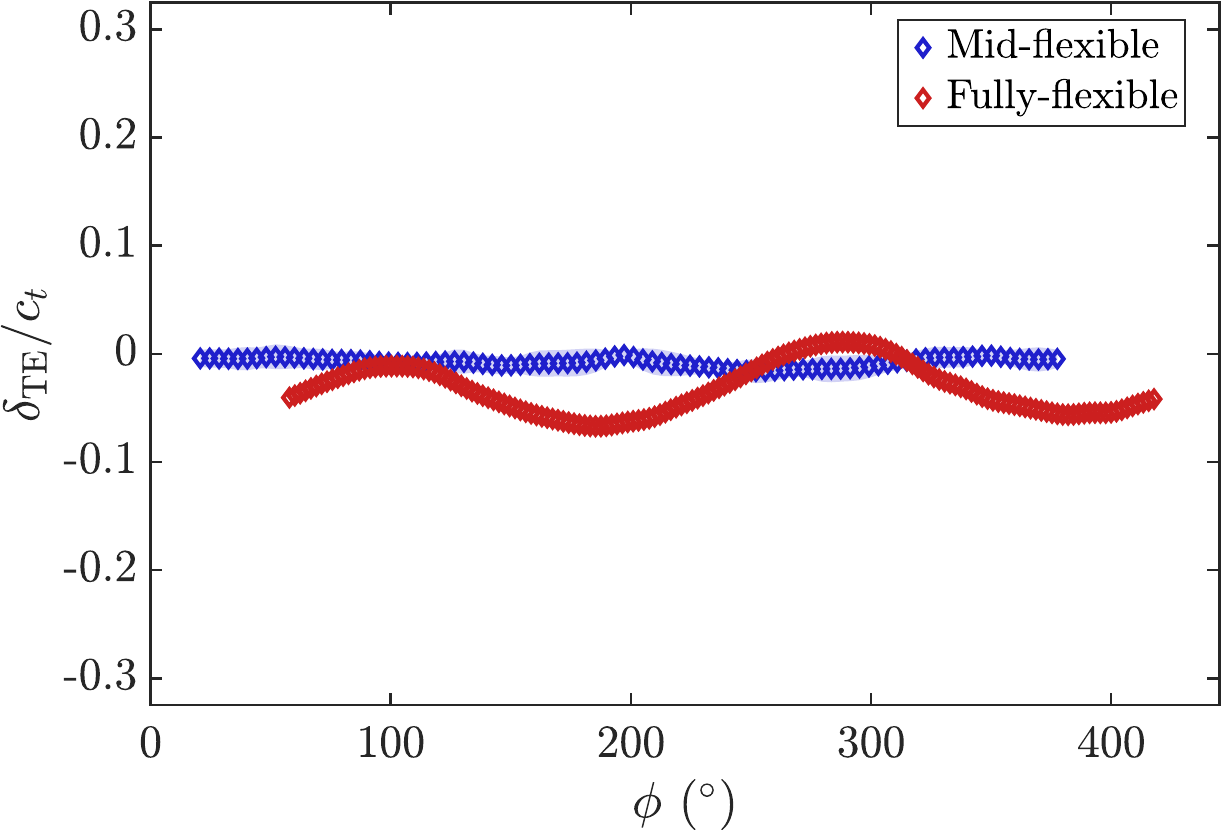}
         \caption{}
         \label{subfig:trackingXOneCondition}
     \end{subfigure}%

     \centering
     \begin{subfigure}{0.5\textwidth}
         \includegraphics[width=0.9\textwidth]{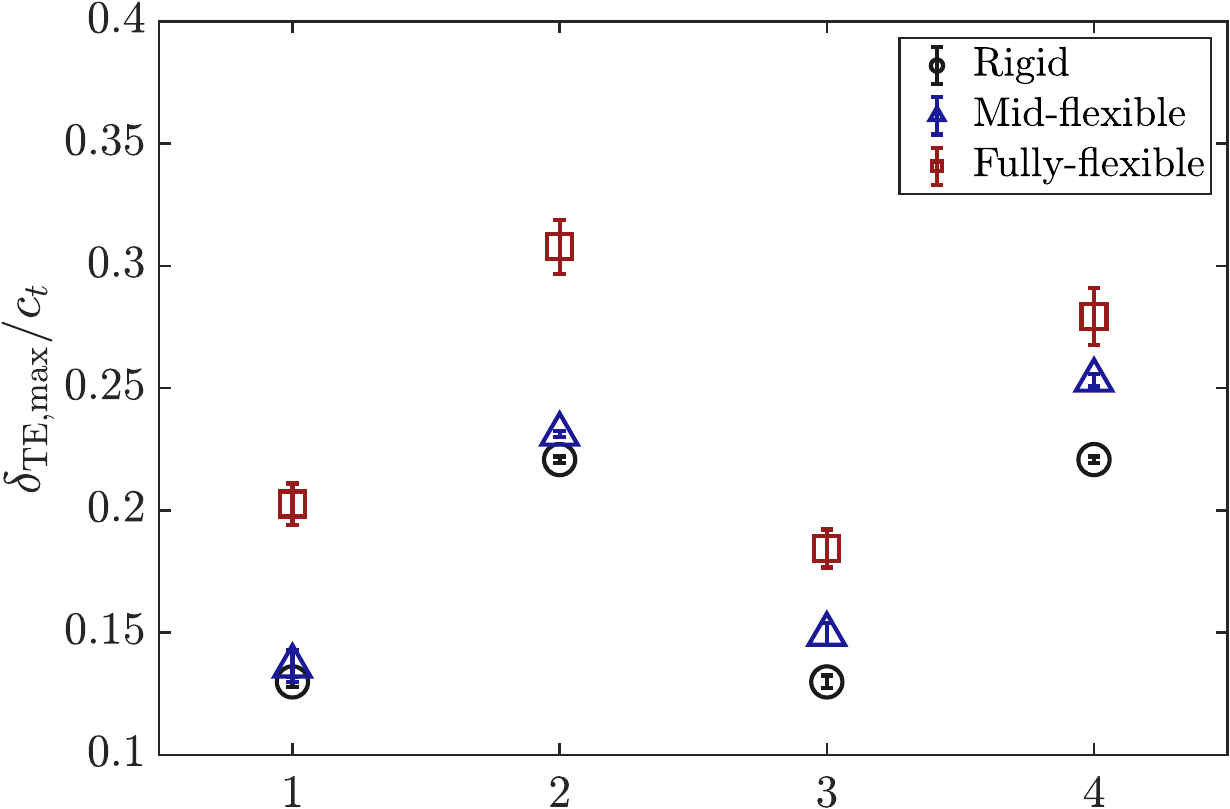}
         \caption{}
         \label{subfig:trackingExtracted}
     \end{subfigure}%
     \hfill
     \begin{subfigure}{0.5\textwidth}
         \centering
         \includegraphics[width=0.9\textwidth]{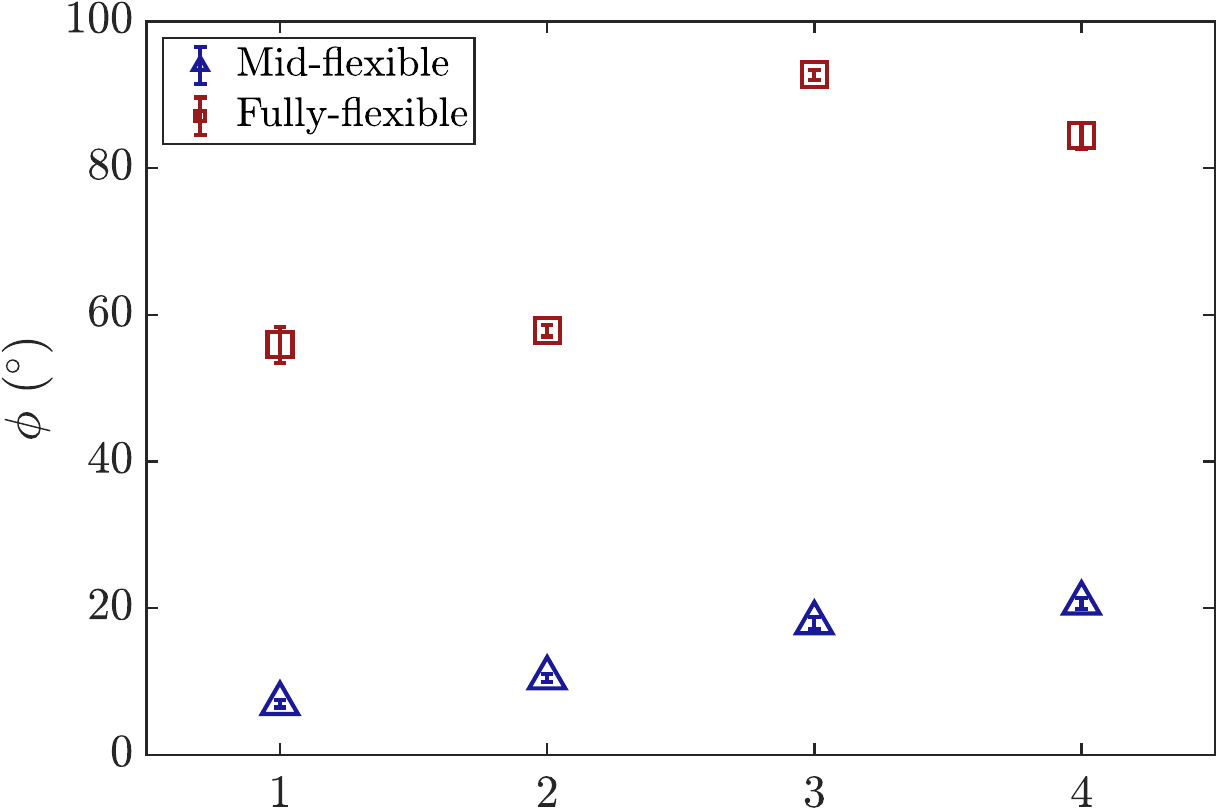}
         \caption{}
         \label{subfig:trackingExtractedPhase}
     \end{subfigure}%
        \caption{Comparison of trailing edge deflection ($\delta_{TE}$) in (a) spanwise and (b) streamwise direction between rigid, mid-flexible and fully-flexible cases at heave frequency and amplitude of $1.25~\mathrm{Hz}$ and $h^*=0.22$, respectively. (c) Maximum trailing edge amplitude (normalized by chord) at different operating conditions, where 1, 2, 3, and 4 corresponds to ($h^*=0.13,f = 0.8~\mathrm{Hz}$), ($h^*=0.22,f = 0.8~\mathrm{Hz}$), ($h^*=0.13,f = 1.25~\mathrm{Hz}$), and ($h^*=0.22,f = 1.25~\mathrm{Hz}$), respectively. (d) Phase lag between the imposed heaving motion and the tail motion for different operating conditions.}
        \label{fig:tracking}
\end{figure}

The fully-flexible foil undergoes the largest trailing-edge excursions and exhibits the greatest phase delay in trailing-edge motion among the three configurations, directly influencing the timing of vortex shedding. Trailing-edge excursions and phase delay extracted from optical tracking measurements are summarized in Fig.~\ref{fig:tracking}, and consistently show higher amplitudes for the fully-flexible foil across all operating conditions. These increased excursion amplitudes and phase differences delay the shedding of trailing-edge vortices, thereby increasing their streamwise separation and enhancing vortex longevity. Notably, the phase delay for the fully-flexible case approaches $90^\circ$, relative to the other rigidity cases. A phase lag of approximately $90^\circ$ between heave and pitch motions in flapping foils has been associated with optimal thrust production and the formation of a reverse von K\'{a}rm\'{a}n vortex street \citep{anderson1998oscillating,raut2024hydrodynamic}. The combined effect of larger trailing-edge excursions and near-optimal phase difference leads to a favorable delay in vortex shedding, which is directly evident in Fig.~\ref{fig:085HzPhaseAveraged}. Coherent positive vorticity structures are already present within the field of view for the rigid and mid-flexible cases (Fig.~\ref{fig:085HzPhaseAveraged}(a,b)), whereas the corresponding vortex for the fully-flexible case has not yet entered the measurement region. A similar delay is observed during the second half of the heave cycle. This increased vortex spacing suppresses destructive vortex interactions, resulting in enhanced jet persistence and consequently higher propulsive efficiency.\\

While the vortex characteristics varied significantly for the fully-flexible case, the timing of vortex shedding for the mid-flexible case closely resembles that of the rigid foil. Although the mid-flexible foil exhibits marginally larger trailing-edge excursions, these deformations are insufficient to produce a qualitative change in the vortex-shedding timing. Consequently, the jet longevity, and therefore the propulsive efficiency, of the mid-flexible case remains comparable to that of the rigid foil at the operating condition shown in the figure. Similar behavior was observed at other operating conditions, including $0.8~\mathrm{Hz},~h^* = 0.13$ and $1.25~\mathrm{Hz},~h^* = 0.13$ (not shown).

\begin{figure}
     \centering
         \includegraphics[width=0.5\textwidth]{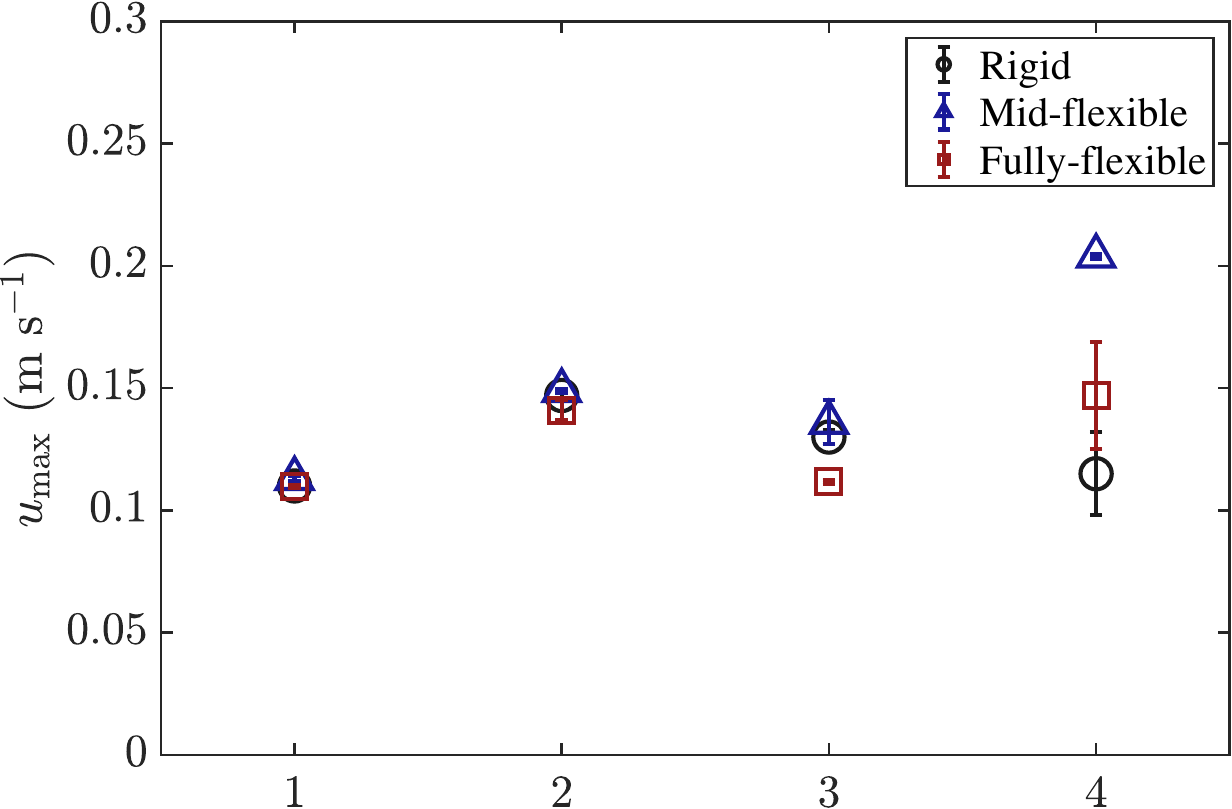}
        
        \caption{Comparison of peak streamwise jet velocity between rigid, mid-flexible and fully-flexible cases. The labels in the x-axis represents different operating conditions where 1, 2, 3, and 4 corresponds to ($h=0.13,f = 0.8~\mathrm{Hz}$), ($h=0.22,f = 0.8~\mathrm{Hz}$), ($h=0.13,f = 1.25~\mathrm{Hz}$), and ($h=0.22,f = 1.25~\mathrm{Hz}$), respectively.}
        \label{fig:umaxExtract}
\end{figure}

Next, we extract the maximum amplitude of the peak velocity ($u_{max}$) from Fig.~\ref{fig:meanAndMaxVelocities} to draw correlations from the thrust characteristics. The results are summarized in Fig. \ref{fig:umaxExtract}. At conditions 1, 2, and 3, the differences in the maximum amplitude of $u_{max}$ are relatively small, similar to the observed thrust characteristics at the same conditions. Since these differences were small, no discernible correlation were observed for these three conditions. However, examining condition 4 reveals that the mid-flexible has substantially higher $u_{max}$ which also corresponds to the higher observed thrust at the same condition with the mid-flexible case. This relationship suggests a positive correlation between the mean thrust and the time-averaged flow characteristics, with substantially stronger jet velocity close to the trailing edge associated with higher thrust production.\\

While the thrust and time-averaged flow characteristics show good correlation, the question still remains as to why the mid-flexible case results in stronger jet and higher thrust when compared to the fully-flexible case, which has higher trailing edge deflection with higher expected streamwise momentum in the wake. To answer this question, the streamwise deflection of the flexible tail is also extracted and plotted in Fig.~\ref{fig:tracking}c, with negative values representing the deflection in the negative streamwise direction with reference to the initial tail location before the motion starts. The tail deflection in the negative streamwise direction is significantly higher for the fully-flexible compared to the mid-flexible case. As a consequence, the momentum imparted to the fluid by the mid-flexible case is mostly concentrated in the streamwise direction contributing to thrust, while that by the fully-flexible case is relatively diffused in the streamwise direction. Thus, the mid-flexible case has higher jet velocity and higher thrust than the fully-flexible case. 

\section{Discussion}\label{sec:discussion}
The trends observed in the present measurements are broadly consistent with prior studies on flexible heaving foils. Heathcote \cite{heathcote_flexible_2004} investigated the influence of flexibility by varying the tail rigidity while holding the flexure location and tail length fixed, whereas the present study varies the flexure location at fixed rigidity. Heathcote reported that foils with intermediate rigidity produced higher thrust over most operating conditions, while the most flexible tails achieved the highest propulsive efficiency. Similar behavior is observed here: modifying the flexure location results in higher thrust for the mid-flexible configuration at higher heave amplitudes and frequencies, while the fully-flexible configuration produces slightly higher thrust at lower amplitudes and frequencies. Across all conditions examined, the fully-flexible case consistently exhibits the highest efficiency. \\

The fully-flexible configuration also exhibits the largest trailing-edge deflections, leading to increased vortex spacing and enhanced jet persistence. This increased wake longevity correlates well with the observed improvements in efficiency, consistent with Heathcote’s findings. In addition, Cleaver \cite{cleaver2014thrust}, in a study of a heaving foil in freestream, showed that maximum thrust occurs at moderate trailing-edge deflections, an observation consistent with Heathcote’s findings and echoed in the present measurements, despite the quiescent operating conditions considered here. Collectively, these results suggest that variations in either tail rigidity or flexure location can produce comparable effects on thrust generation in quiescent heaving foils, primarily through their influence on trailing-edge kinematics.\\

\begin{figure}[h!]
  \centerline{\includegraphics[width=0.45\columnwidth]{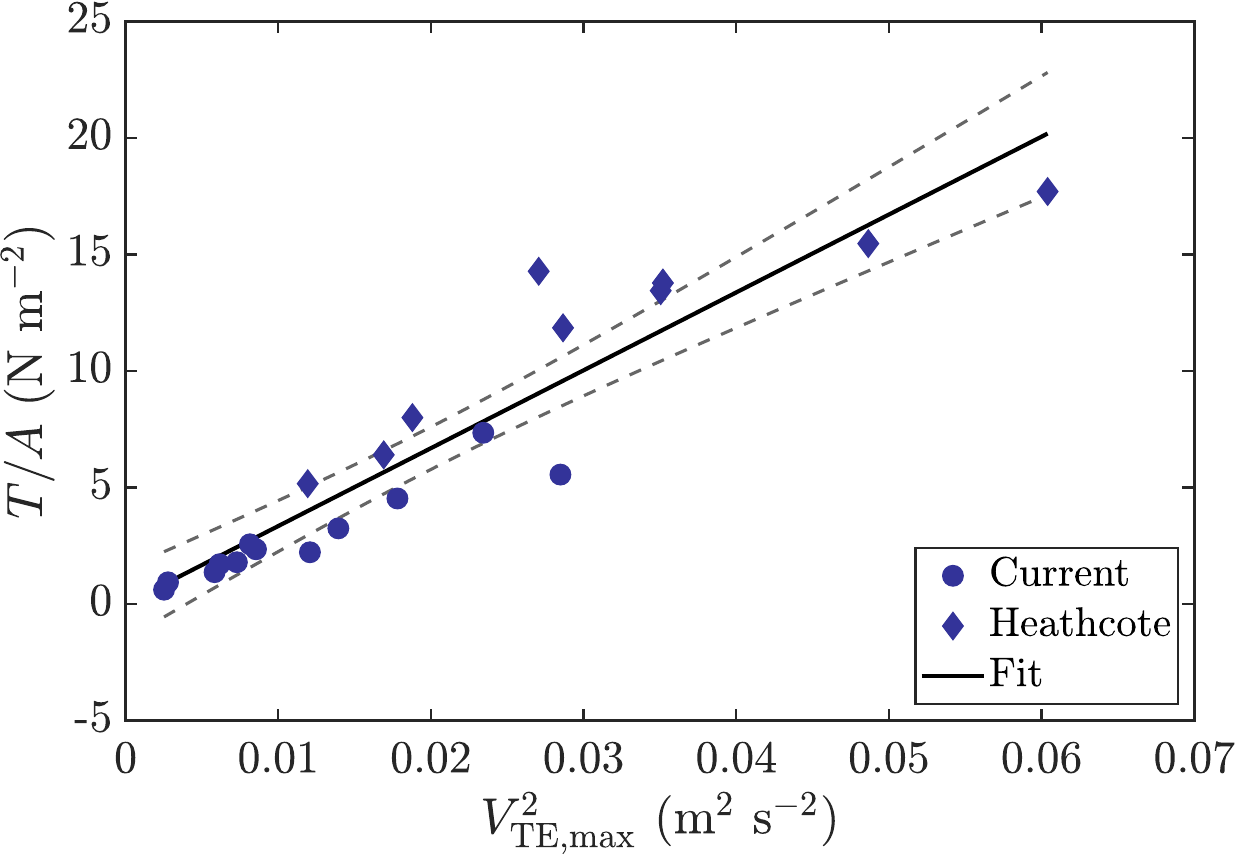}}
  \caption{Thrust per unit area as a function of the square of the maximum trailing-edge velocity. The solid line denotes a linear fit to the present data and to measurements from Heathcote~\citep{heathcote_flexible_2004}.}
\label{fig:fitAnalysis}
\end{figure}

This interpretation is consistent with the observations of David \cite{david2017thrust}, who studied a pitching foil with a flexible tail in freestream and introduced a nondimensional rigidity parameter,
\begin{equation}
R^* = \frac{EI}{0.5 \rho U_\infty^2 c_f^3},
\end{equation}
where $EI$ is the flexural rigidity, $\rho$ is the fluid density, $U_\infty$ is the freestream velocity, and $c_f$ is the flexible tail length. The parameter $R^*$ incorporates both material and geometric effects for a given reduced frequency, such that changes in either rigidity or flexure location yield similar aerodynamic responses when $R^*$ is held constant. However, the applicability of this scaling has not been examined for heaving foils in quiescent conditions. In the absence of a freestream velocity, an alternative velocity scale is required, and the development of appropriate scaling laws for quiescent flows remains an open problem that warrants further investigation.\\

One candidate scaling parameter is the maximum trailing-edge velocity, which has been identified as a key determinant of thrust production. Previous studies have shown that the dimensional thrust scales linearly with the square of the maximum trailing-edge velocity \citep{floryan2020swimmers}. Here, we examine whether this relationship extends to flexible heaving foils in quiescent conditions. To account for both rigidity and flexure location, data from Heathcote \citep{heathcote_flexible_2004} is also included in Fig.~\ref{fig:fitAnalysis}, with thrust normalized by foil span and overall chord to account for geometric differences between the two studies. The maximum trailing-edge velocity is defined as $V_{TE_{\max}} = 2 \pi a_{TE} f$, where $a_{TE}$ is the maximum trailing-edge displacement and $f$ is the heave frequency.\\

The combined data exhibit an approximately linear dependence on $V_{TE_{\max}}^2$ at lower velocities. At higher trailing-edge velocities, increased scatter is observed, suggesting the onset of flow separation effects. It is hypothesized that further increase in trailing-edge velocity would lead to greater deviations from this scaling, particularly as leading-edge vortices begin to dominate thrust production \cite{raut2024hydrodynamic}. These observations motivate future studies at higher amplitudes and frequencies. Nevertheless, the present results demonstrate that the maximum trailing-edge velocity serves as an effective scaling parameter for thrust production in flexible heaving foils under quiescent conditions, particularly at lower heave amplitudes and frequencies.

\section{Conclusions}\label{sec:conclusion}
The influence of flexible-tail length on the performance of a purely heaving foil in quiescent flow was investigated by systematically varying the flexure location while maintaining constant total chord length ($c_t = 0.077~\mathrm{m}$) and flexural rigidity ($EI = 1.57 \times 10^{-5}~\mathrm{Nm^2}$). Two compliant configurations were considered: a \emph{mid-flexible} tail with a flexible length of $0.022~\mathrm{m}$ and a \emph{fully-flexible} tail with a flexible length of $0.044~\mathrm{m}$. A rigid-tail configuration of identical thickness served as the baseline. Force and flow-field measurements were performed at heave frequencies of $0.8~\mathrm{Hz}$ and $1.25~\mathrm{Hz}$ and non-dimensional heave amplitudes of $h^* = 0.13$ and $0.22$ to examine how the spatial distribution of flexibility influences thrust generation, wake dynamics, and propulsive efficiency.\\

The fully-flexible configuration consistently achieved higher propulsive efficiency than both the rigid and mid-flexible cases across all operating conditions. Time-averaged flow fields revealed enhanced jet persistence and a more coherent downstream momentum jet for the fully-flexible tail. Phase-averaged analysis further showed that the fully-flexible configuration increased the streamwise spacing between successive vortices by delaying trailing-edge vortex shedding. The increased vortex spacing allowed vortical structures to convect farther downstream before interacting, thereby sustaining jet coherence and contributing to the observed improvement in efficiency. These results demonstrate that increasing flexible length can fundamentally reorganize the wake and strengthen the persistence of the momentum jet. In contrast, the mid-flexible configuration generated the highest thrust at the largest heave amplitude and frequency tested. Although the mid-flexible tail underwent smaller overall deformations than the fully-flexible case, it produced a stronger near-wake jet with higher momentum flux, consistent with the measured increase in thrust. Kinematic measurements indicate that the fully-flexible tail experiences larger negative streamwise deflections at high forcing amplitudes. These deformations likely reduce the effective transfer of streamwise momentum to the fluid, thereby limiting thrust production despite the improved wake coherence. The results therefore suggest that excessive compliance can become detrimental to thrust generation when large streamwise deformations reduce effective momentum injection into the wake.\\

Overall, the findings demonstrate that the spatial distribution of flexibility governs the trade-off between thrust generation and propulsive efficiency by altering vortex dynamics, wake organization, and the coupling between structural deformation and vortex shedding. Increasing flexible length promotes wake coherence and enhances efficiency, whereas an intermediate level of compliance can strengthen near-wake momentum injection and maximize thrust under high forcing conditions. Beyond the specific configuration examined here, the present study advances the fluid mechanics understanding of how distributed flexibility modifies vortex formation, jet development, and momentum transfer in unsteady propulsive systems. The results provide mechanistic insight into the fluid--structure interaction pathways through which flexibility distribution alters hydrodynamic performance, thereby contributing to the broader design framework for bio-inspired and engineered propulsion systems. These findings are directly relevant to the development of efficient wave-powered marine vehicles, flexible propulsors, autonomous underwater systems, and energy-harvesting devices, where tailoring the spatial distribution of compliance can be used to optimize competing performance objectives such as thrust production, efficiency, and wake stability.\\

Future work should examine the coupled effects of flexible length, flexural rigidity, and forcing kinematics over a broader range of Strouhal numbers and operating conditions. Extension to combined heave--pitch motions and operation in currents and wave environments will be important for assessing relevance to realistic marine propulsion and wave-energy-harvesting applications. Adaptive or spatially varying stiffness distributions should also be explored as a means of achieving real-time performance optimization in autonomous marine platforms where sea-state variability demands propulsors capable of switching between thrust- and efficiency-optimized operating regimes.

\section*{Acknowledgments}

\noindent The preliminary investigation of this work was conducted as part of the graduate course on Experimental Methods in Fluid Mechanics (NA599) taught by Prof. Anchal Sareen in the Department of Naval Architecture and Marine Engineering at the University of Michigan, Ann Arbor. The authors acknowledge Madeline Marusich, Riley McKenna, Yolanda Ming, Neel Karani and Eamonn Reilly for their contributions to the conceptual development of the study and for conducting the preliminary experiments. The Office of Naval Research under Grant No.~N00014-24-1-2013 is acknowledged for providing the salary support for Lokesh Silwal and the Department of Naval Architecture and Marine Engineering is acknowledged for providing funds to run the class.

\section*{Author Declarations}
\noindent \textbf{Conflict of Interest}\\
 The authors have no conflicts to disclose.

\section*{Author Contributions}

\textbf{Lokesh Silwal}: Formal analysis (equal); Investigation (equal); Methodology (equal); Validation (equal); Writing - original draft (equal).
\textbf{Neel Karani}: Formal analysis (equal); Investigation (equal); Methodology (equal); Validation (equal); Writing - review \& editing.
\textbf{Anchal Sareen}: Formal analysis (equal); Investigation (equal); Methodology (equal); Validation (equal); Funding acquisition (lead); Resources (lead); Supervision (lead); Writing - review \& editing (lead). 

\section*{Data Availability Statement}
Data supporting the findings of this study are available from the corresponding author on a reasonable request.

\section{Appendix}\label{appA}
\appendix

\begin{figure}[h!]\centering
  \begin{subfigure}{0.5\textwidth}
         \centering
         \includegraphics[width=0.9\textwidth]{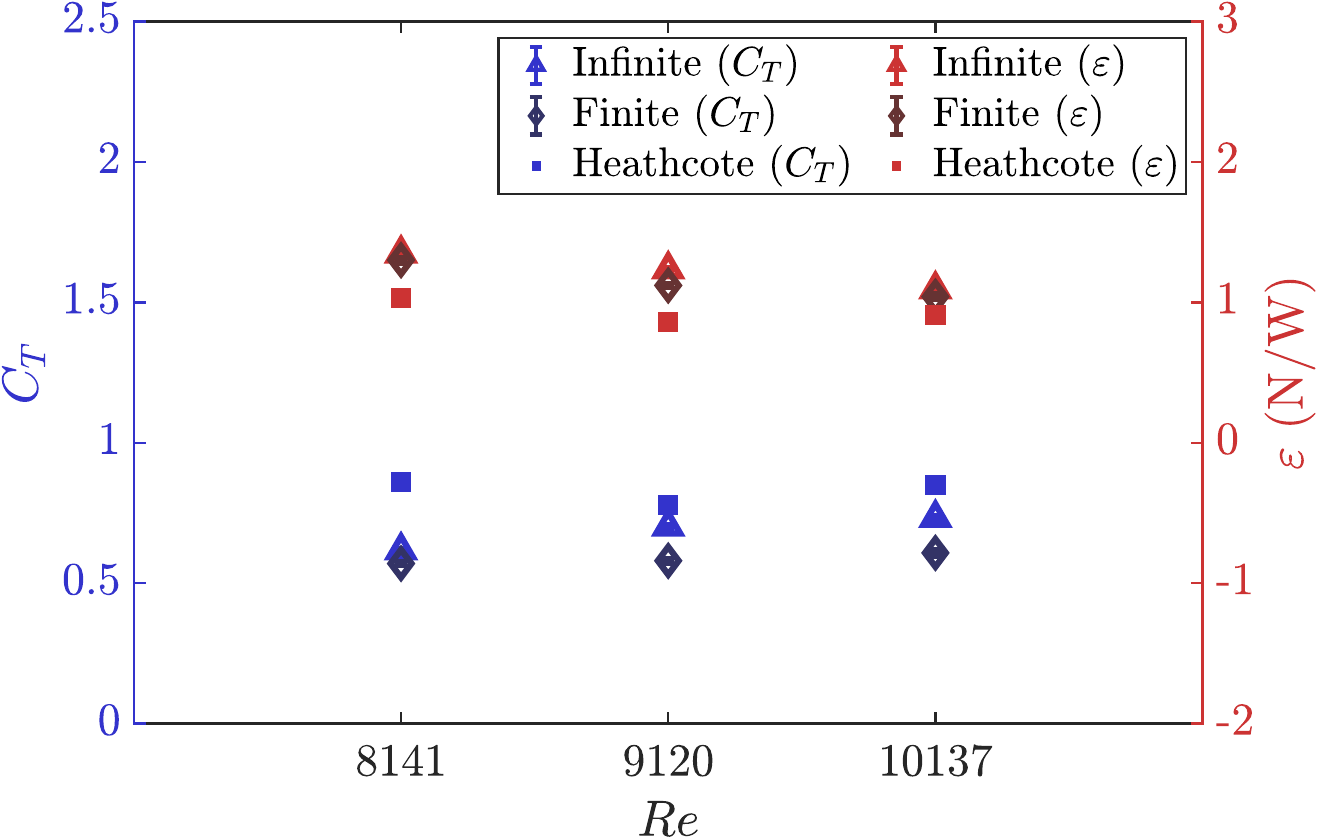}
         \caption{}
         \label{subfig:validationInfiniteCoeffThrust}
     \end{subfigure}%
     \hfill
     \begin{subfigure}{0.5\textwidth}
         \centering
         \includegraphics[width=0.9\textwidth]{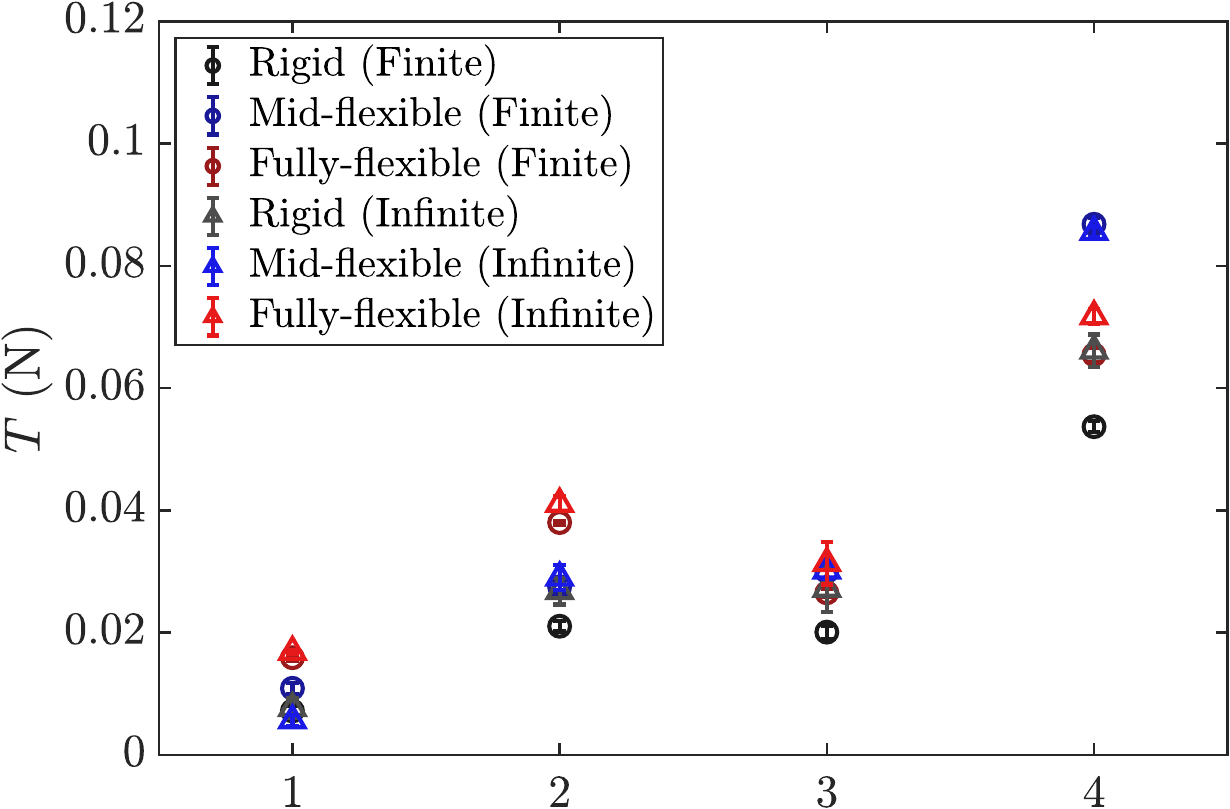}
         \caption{}
         \label{subfig:validationInfiniteThrust}
     \end{subfigure}%
  \caption{(a) Variation of thrust coefficient for a fully rigid foil with finite and infinite end conditions in comparison with Heathcote's measurements. (b) Comparison of thrust between finite and infinite conditions across the three types of the foils tested. }
\label{fig:forcesFiniteAndInfinite}
\end{figure}

In this section, the performance of the finite-span foil is compared with the infinite-span condition approximated using end plates at both ends of the foil. As shown in Fig~\ref{fig:forcesFiniteAndInfinite}b, the thrust produced is largely unchanged between the finite- and infinite-span configurations, except for the fully rigid case. For the fully rigid foil, the finite-span configuration yields lower thrust, particularly at the highest frequency and heave amplitude considered. 

\begin{figure}[h!]\centering
  \begin{subfigure}{0.5\textwidth}
         \includegraphics[width=0.9\textwidth]{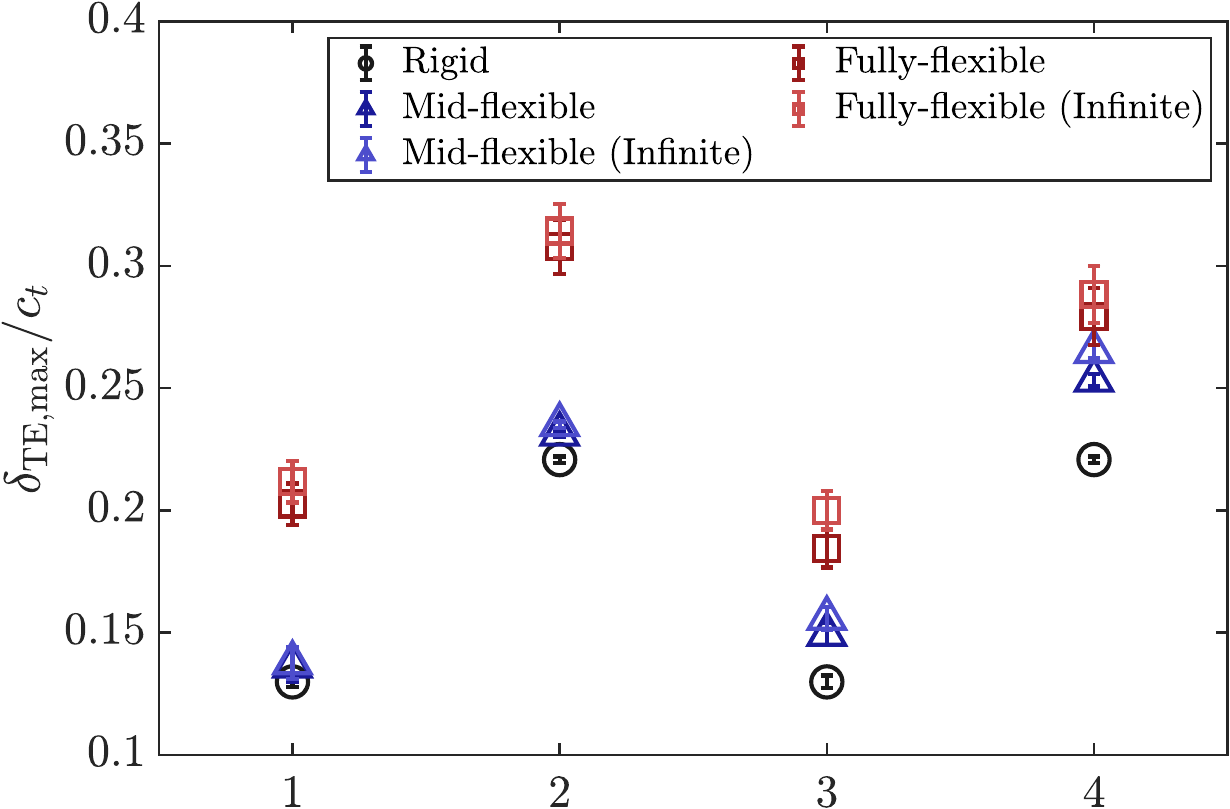}
         \caption{}
         \label{subfig:trackingExtractedInfinite}
     \end{subfigure}%
     \hfill
     \begin{subfigure}{0.5\textwidth}
         \centering
         \includegraphics[width=0.9\textwidth]{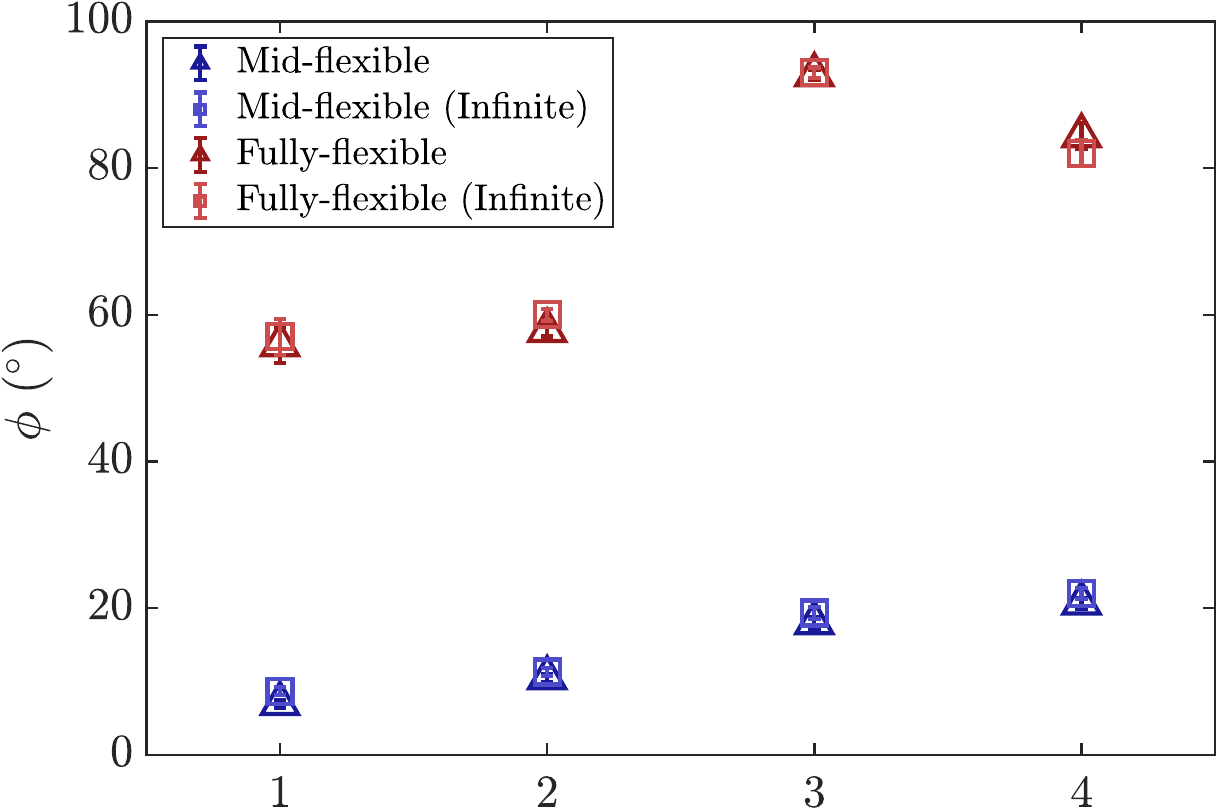}
         \caption{}
         \label{subfig:trackingExtractedPhaseInfinite}
     \end{subfigure}%
  \caption{(a) Maximum trailing edge amplitude (normalized by chord) of the infinite and finite foil at different operating conditions, where 1, 2, 3, and 4 corresponds to ($h^*=0.13,f = 0.8~\mathrm{Hz}$), ($h^*=0.22,f = 0.8~\mathrm{Hz}$), ($h^*=0.13,f = 1.25~\mathrm{Hz}$), and ($h^*=0.22,f = 1.25~\mathrm{Hz}$), respectively. (b) Phase lag between the imposed heaving motion and the tail motion for different operating conditions of the infinite and finite foil.}
\label{fig:trackingFiniteAndInfinite}
\end{figure}

Figure~\ref{fig:trackingFiniteAndInfinite} further shows that the foil kinematics, including the maximum trailing-edge excursion and the phase lag between pitch and heave, remain essentially unchanged between the finite- and infinite-span cases.

\bibliographystyle{elsarticle-num} 
\bibliography{aipsamp}
\vspace{0.5 in}

\end{document}